\documentclass[journal,final]{IEEEtran}
\usepackage[dvipsnames]{xcolor}
\usepackage{gensymb}
\usepackage{cite}
\usepackage{array}
\usepackage{caption}
\usepackage{subcaption}
\usepackage{hyperref}
\usepackage{booktabs} % To thicken table lines

\newcommand{\rev}[1]{\textcolor{black}{#1}}

\ifCLASSINFOpdf
   \usepackage{graphicx}
\else
\fi
\usepackage{amsmath}

\begin{document}
% Titles are generally capitalized except for words such as a, an, and, as, at, but, by, for, in, nor, of, on, or, the, to and up, which are usually not capitalized unless they are the first or last word of the title.
% Linebreaks \\ can be used within to get better formatting as desired. Do not put math or special symbols in the title.
\title{Wide-Area Land Cover Mapping with Sentinel-1 Imagery using Deep Learning Semantic Segmentation Models}
% note positions of commas and nonbreaking spaces ( ~ ) LaTeX will not break a structure at a ~ so this keeps an author's name from being broken across two lines.
% a separate \thanks must be used for each paragraph as LaTeX2e's \thanks was not built to handle multiple paragraphs
%

\author{Sanja~\v{S}\'{c}epanovi\'{c},
        Oleg~Antropov,~\IEEEmembership{Member,~IEEE,}
        Pekka~Laurila,
        Yrjö~Rauste,
        Vladimir~Ignatenko,
        and~Jaan~Praks,~\IEEEmembership{Member,~IEEE}% <-this % stops a space
\thanks{Sanja~\v{S}\'{c}epanovi\'{c}, Pekka Laurila and Vladimir Ignatenko were with ICEYE Oy, Espoo, Finland e-mail: name.surname@iceye.fi}% <-this % stops a space
%\thanks{Sanja~\v{S}\'{c}epanovi\'{c} was also with Nokia Bell Labs, email: sanja.scepanovic@nokia-bell-labs.com.}% <-this % stops a space
\thanks{Oleg Antropov was with VTT Technical Research Centre of Finland, Espoo, Finland, email: oleg.antropov@vtt.fi.}% <-this % stops a space
\thanks{Jaan Praks was with Aalto University, Espoo, Finland, email: jaan.praks@aalto.fi.}% <-this % stops a space
\thanks{Manuscript received 28.9.2020 }}

\maketitle

% As a general rule, do not put math, special symbols or citations in the abstract or keywords.
\begin{abstract}
Land cover mapping is essential for monitoring the environment and understanding the effects of human activities on it. The automatic approaches to land cover mapping (i.e., image segmentation) mostly used traditional machine learning that requires heuristic feature design. On the natural images, deep learning has outperformed traditional machine learning approaches on a range of tasks, including the image segmentation. On remote sensing images, recent studies are demonstrating successful application of specific deep learning models or their adaptations to particular small-scale land cover mapping tasks (e.g., to classify wetland complexes). However, it is not readily clear which of the existing state-of-the-art models for natural images are the best candidates to be taken for the particular remote sensing task and data. 

In this study, we answer that question for mapping the fundamental land cover classes using the satellite imaging radar data. We took ESA Sentinel-1 C-band SAR images available at no cost to users as representative data. CORINE land cover map produced by the Finnish Environment Institute was used as a reference, and the models were trained to distinguish between the 5 major CORINE based classes.  We selected seven among the state-of-the-art semantic segmentation models so that they cover a diverse set of approaches: U-Net, DeepLabV3+,  PSPNet, BiSeNet, SegNet, FC-DenseNet, and FRRN-B.  These models were pre-trained on the ImageNet dataset and further fine-tuned in this study.  Specifically, we used ESA Sentinel-1 scenes acquired during the whole summer season of 2018 in Finland, which are representative of the land cover in the country. 

\rev{Upon evaluation and benchmarking, all the models demonstrated solid performance with overall accuracy between 87.9\% and 93.1\%, with good to very good agreement (kappa statistic between 0.75 and 0.86). The two best models were FC-DenseNet (Fully Convolutional DenseNets) and SegNet  (Encoder-Decoder-Skip), with the latter having much smaller inference time. Overall, our results indicate that the semantic segmentation models are suitable for efficient wide-area mapping using satellite SAR imagery. Our results also provide baseline accuracy against which the newly proposed models should be evaluated and suggest the DenseNet- and SegNet-based models are the first candidates for this task.}\footnote{This work has been submitted to the IEEE for possible publication. Copyright may be
	transferred without notice, after which this version may no longer be accessible.} 
\end{abstract}


% \fi
\IEEEpeerreviewmaketitle

\section{Introduction}

\IEEEPARstart{M}{apping } of land cover and its change has a critical role in the characterization of the current state of the environment. The changes in land cover can be due either to human activities as well as caused by climate changes on a regional scale. The land cover, on the other hand, affects climate through water and energy exchange with the atmosphere and by changing carbon balance. Because of this, land cover belongs to the Essential Climate Variables \cite{bojinski2014}.  Hence, timely assessment of land cover and its change is one of the most important applications in satellite remote sensing.
Thematic maps are needed annually for various purposes in medium resolution (circa 250 m) with less than 15\% measurement uncertainty and in high resolution (10-30 m) with less than 5\% uncertainty. 

CORINE Land Cover (CLC) is a notable example of a consistent Pan-European land cover mapping initiative \cite{buttner2004corine,bossard2000corine} coordinated by the European Environment Agency (EEA).\footnote{https://land.copernicus.eu/pan-european/corine-land-cover} CORINE stands for \textit{coordination of information on the environment}. It is an on-going long-term effort providing most harmonized classification land cover data in Europe with updates approximately every 4 years. The CORINE maps are an important source of land cover information suitable for operational purposes also for various customer groups in Europe. It has altogether 44 classes, though many of them are not strictly ecological classes but rather land use classes. On the continental scale, CORINE provides a harmonized map with 25 ha minimum mapping unit (MMU) for areal phenomena, and a minimum width of 100 m for linear phenomena \cite{buttner2014corine}. National land cover maps in the CORINE framework can exhibit smaller mapping units. In Finland, the latest revision of CORINE land cover map at the time of this study was year 2018 version produced by the Finnish Environment Institute. The map has an MMU of $20m \times 20m$ and was produced by a combined automated and manual interpretation of the high-resolution satellite optical data followed by the data integration with existing basic map layers \cite{torma2015}. 

The state-of-the-art approaches used for land cover mapping mainly rely on the satellite optical imagery. The key role is played by the Landsat imagery often augmented by the MODIS or SPOT-5 imagery \cite{chen2015global,almeida2016high,homer2015completion}.
Other sources of information employed for land cover mapping include Digital Elevation Models (DEM) and very high-resolution imagery \cite{zhao2016detailed}. When it comes to the large-scale and multitemporal land cover mapping, a more recent optical imagery source is Copernicus Sentinel-2. With a revisit of 5 days, it has become another key data source \cite{griffiths2019intra}. 

International programs, such as the European Space Agency's (ESA's) Copernicus \cite{torres2012} behind the Sentinel satellites are taking significant efforts to make Earth Observation (EO) data freely available for commercial and non-commercial purposes. The Copernicus programme is a multi-billion investment by the EU and ESA aiming to provide essential services based on accurate and timely data from satellites. Its main goals are to improve the ways of managing the environment, to help mitigate the effects of climate change, and enable the creation of new applications and services, such as for environmental monitoring and urban development.

The provision of free satellite data for mapping in the framework of such programs also enables application of methods that could not be used earlier because they require vast and representative datasets for training, for example \textit{deep learning}. In recent years, deep learning has brought about several breakthroughs in the pattern recognition and computer vision \cite{lecun1998gradient,krizhevsky2012imagenet,Simonyan2014VeryDC}. 
The success of the deep learning models can be attributed to both their \textit{deep multilayer structure} creating nonlinear functions and, hence, allowing extraction of hierarchical sets of features from the data, and to their \textit{end-to-end training scheme} allowing for simultaneous learning of the features from the raw input and predicting the task at hand. In this way, the heuristic feature design is removed. This is advantageous compared to the traditional machine learning methods (e.g., support vector machine (SVM) and random forest (RF)), which require a multistage feature engineering procedure. In deep learning, such a procedure is replaced with a simple end-to-end deep learning workflow. One of the key requirements for successful application of deep learning methods is a large amount of data available from which the model can automatically learn the representative features for the prediction task \cite{goodfellow2016deep}. The availability of open satellite imagery, such as from Copernicus, offers just that.

The land cover mapping systems based solely on optical imagery suffer from issues with cloud cover and weather conditions, especially in the tropical areas, and with a lack of illumination in the polar regions. Among the free satellite data offered by the Copernicus programme are \textit{synthetic aperture radar (SAR)} images from the Sentinel-1 satellites. SAR is an active radar imaging technique that does not require illumination and is not hampered by cloud-cover due to penetration of microwave radiation through clouds. The utilisation of SAR imagery, hence, would allow mapping such challenging regions and increasing the mapping frequency in the orchestrated efforts like CORINE. One of the significant issues previously was the absence of timely and consistent high-resolution wide-area SAR coverage. With the advent of Copernicus Sentinel-1 satellites, operational use of imaging radar data becomes feasible for consistent wide-area mapping. The first Copernicus Sentinel-1 mission was launched in April 2014. Firstly, Sentinel-1A alone was capable of providing C-band SAR data in up to four imaging modes with a revisit time of 12 days. Once Sentinel-1B was launched in 2016 the revisit time has reduced to 6 days \cite{torres2012}. 

We studied wide-area SAR-based land cover mapping by methodologically combining the two discussed recent advances: the improved methods for large-scale image processing using deep learning and the availability of SAR imagery from the Sentinel-1 satellites.  

%%%%%%%%%%%%%%%%%%%%%%%%%%%%%%%%%%%%%%%%%%%%%%%%%%%%%%%%%%%%%%%%%%%%%%%%%%%%%%%%%%%%%%%%%%%%%%%%%%%%%%%%%%%%%%%%%%%%%%%%%%%%%%
\subsection{Land Cover Mapping with SAR Imagery} 
While using optical satellite data is still a mainstream in land cover and land cover change mapping \cite{cohen2004landsat, goetz2009carbonmapping, atzberger2013agro, hame2013, torma2015}, SAR data has been getting more attention as more suitable sensors appear. To date, several studies have investigated the suitability of SAR for land cover mapping, focusing primarily \rev{on} L-band, C-band, and X-band polarimetric \cite{antropov2014, lonnqvist2010} multitemporal and multi-frequency SAR \cite{waske2009lcsar} \cite{bruzzone2004class}, as well as, \rev{on} the combined use of SAR and optical data \cite{ullmann2014tundra, clerici2017fusion,castaneda2009land,ban2010fusion, laurin2013optical}. 

Independently of the imagery used, the majority of land cover mapping methods so far are based on traditional supervised classification techniques \cite{khatami2016meta}. Widely used classifiers are support vector machines (SVM), decision trees, random forests (RF), and maximum likelihood classifiers (MLC) \cite{zhao2016detailed,almeida2016high,waske2009classifier,khatami2016meta}. However, extracting a large number of features needed for classification, i.e., the \textit{feature engineering process}, is time intensive, and requires lots of expert work in developing \rev{and} fine-tuning classification approaches. This limits the applications of the traditional supervised classification methods on a large scale. 

Backscattered microwave radiation is composed \rev{of} multiple fundamental scattering mechanisms determined by the vegetation water content, surface roughness, soil moisture,  horizontal and vertical  structure of the scatterers, as well as imaging geometry during the datatake. Accordingly, a considerable number of classes can be differentiated in SAR images \cite{balzter2015,antropov2014}. However, majority of SAR classification algorithms use fixed SAR observables (e.g., polarimetric features) to infer specific land cover classes, despite the large temporal, seasonal and environmental variability between different geographical sites. This leads to a lack of generalisation capability and a need to use extensive and representative reference data and SAR data. The latter means the need to account for not only all variation of SAR signatures for a specific class but also the need to consider seasonal effects, as changes in moisture of soil and vegetation, as well as frozen state of land \cite{park2015freeze} that strongly affect SAR backscatter. On the other hand, when using multitemporal approaches, such seasonal variation can be used as an effective discriminator among different land cover classes.

When exclusively using SAR data for land cover mapping, reported accuracy often turns out to be relatively low for operational land cover mapping and change monitoring. Methodologically, reported solutions utilized supervised approaches, linking SAR observables and class labels to pixels, superpixels or objects in parametric or nonparametric manner \cite{antropov2014,lonnqvist2010,dobson1996, balzter2015, hame2013, sirro2018, alban2018, longepe2015, esch2011, cable2014, niu2013r2, evans2010, lumsdon2005}.

However, tackling relatively large number of classes was considered only in several studies, often with relatively low reported accuracies. For instance, in \cite{da2008land} it was found that P-band PolSAR imagery was unsatisfactory for mapping more than five classes with the iterated conditional mode (ICM) contextual classifier applied to several polarimetric parameters. They achieved a Kappa value of 76.8\% when mapping four classes.
Classification performance of the L-band ALOS PALSAR and C-band RADARSAT-2 images was compared in the moist tropics \cite{li2012tropical}. L-band provided 72.2\% classification accuracy for a coarse land cover classification system and C-band only 54.7\%.In a similar study in Lao PDR, ALOS PALSAR data were found to be mostly useful as a back-up option to optical ALOS AVNIR data\cite{hame2013}. Multitemporal Radarsat‐1 data with HH polarization and ENVISAT ASAR data with VV polarization (both C-band) were studied for classification of five land cover classes in Korea with moderate accuracy \cite{park2008cband}. Waske \textit{et al.} \cite{waske2009classifier} applied boosted decision tree and random forests to multi-temporal C-band SAR data reaching accuracy up to 84\%.
Several studies \cite{lonnqvist2010}, \cite{antropov2014} investigated specifically SAR suitability for the boreal zone, with reported accuracy up to 83\% depending on the classification technique (maximum likelihood, probabilistic neural networks, etc.) when five super-classes (based on CORINE data) were used. 

The potential of Sentinel-1 imagery for CORINE-type thematic mapping was assessed in a study that used Sentinel-1A data for mapping class composition in Thuringia \cite{balzter2015}. Long-time series of Sentinel-1 SAR data are considered especially suitable for crop type mapping \cite{tomppo2019, nguyen2016mapping,veloso2017understanding,satalino2013c}, with increased number of studies attempting land cover mapping in general \cite{vicente2018, ge2019}. 

Moreover, as Sentinel-1 data are presently the only free source of SAR data routinely available for wide-area mapping at no cost for users, it seems the best candidate data for development and testing of improved classification approaches. Previous studies indicate a necessity for developing and testing new methodological approaches that can be effectively applied to a large-scale and deal with the variability of SAR observables concerning ecological land cover classes. We suggest adopting state-of-the-art deep learning approaches for this purpose.

\subsection{Deep Learning in Remote Sensing}
The advances in the deep learning techniques for computer vision, in particular, Convolutional Neural Networks (CNNs) \cite{lecun1998gradient,lecun1995convolutional}, have led to the application of deep learning in several domains that rely on computer vision. Examples are self-driving cars, image search engines, medical diagnostics, and augmented reality. \rev{Deep learning approaches are becoming extensively applied in the remote sensing domain, as well.}

Zhu \textit{et al.} \cite{zhu2017deep} provide a discussion on the specificities of remote sensing imagery (compared to ordinary RGB images) that result in specific deep learning challenges in this area. For example, remote sensing data are georeferenced, often multi-modal, with particular imaging geometries, there are interpretation difficulties, and the ground-truth or labelled data needed for deep learning is still often lacking. Additionally, most of the state-of-the-art CNNs are developed for three-channel input images (i.e., RGB) and so certain adaptations are needed to apply them on the remote sensing data \cite{mahdianpari2018very}. 
 
Nevertheless, several research papers tackling remote sensing imagery with deep learning techniques were published in recent years. Zhang \textit{et al.} \cite{zhang2016deep} review the field and find 
 applications to image preprocessing \cite{zhang2014l_}, target recognition \cite{chen2013aircraft,chen2014vehicle}, classification \cite{doi:10.1080/01431161.2015.1055607,wang2015deep,tuia2015multiclass}, and semantic feature extraction and scene understanding \cite{hu2015transferring,7301382,luus2015multiview,zhang2016scene}. The deep learning approaches are found to outperform standard methods applied up to several years ago, i.e., SVMs and RFs \cite{7153200,kussul2017deep}. 
 
When it comes to deep learning for \textit{land cover} or \textit{land use mapping}, applications have been limited to optical satellite \cite{mahdianpari2018very,chen2014deep,mahdianpari2018very,wang2015deep} or aerial  \cite{wu2018automatic} imagery, and hyperspectral imagery \cite{tuia2015multiclass,chen2014deep} owing to the similarity of these images to ordinary RGB images studied in computer vision \cite{mahdianpari2018very}.

When it comes to SAR images, Zhang \textit{et al.} \cite{zhang2016deep} found that there is already a significant success in applying deep learning techniques for object detection and scene understanding. However, for classification on SAR data, applications are scarce and advances are yet to be achieved \cite{zhang2016deep}. Published research includes deep learning for crop types mapping using combined optical and SAR imagery \cite{kussul2017deep}, as well as the use of SAR images exclusively \cite{duan2017sar}.  
However, those methods applied deep learning only to some part of the task at hand and not in an end-to-end fashion. Wang \textit{et al.} \cite{wang2015deep}, for instance, just used deep neural networks for merging over-segmented elements, which are produced using traditional segmentation approaches. Similarly, Tuia \textit{et al.} \cite{tuia2015multiclass} applied deep learning to extract hierarchical features, which they further fed into a multiclass logistic classifier. Duan \textit{et al.} \cite{duan2017sar} used first unsupervised deep learning and then continued with a couple of supervised labelling tasks. Chen \textit{et al.} \cite{chen2014deep} applied a deep learning technique (stacked autoencoders) to discover the features, but then they still used traditional machine learning (SVM, logistic regression) for the image segmentation. Unlike those methods, we applied the deep learning in an \textit{end-to-end fashion}, i.e., from supervised feature extraction to the land class prediction. This makes our approach more flexible, robust and, adaptable to the SAR data from new regions, as well as more efficient.

When it comes to the \textit{end-to-end} approaches for SAR classification, there are several studies where the focus was on a small area and on a specific land cover mapping task. For instance, Mohammadimanesh \textit{et al.} \cite{mohammadimanesh2019} used fully polarimetric SAR (PolSAR) imagery from RADARSAT-2 to classify wetland complexes, for which they have developed a specifically tailored semantic segmentation model.  However, the authors have tackled a small test area (around $10km \times 10km$) and have not explored how their model generalizes to other types of areas. Similarly, Wang \textit{et al.} \cite{wang2018multi} adapted existing CNN models into a fixed-feature-size CNN that they have evaluated on a \textit{small scale} RADARSAT-2 or AIRSAR (i.e., airborne SAR data). In both cases, they have used more advanced fully polarimetric SAR imagery at better resolution as opposed to Sentinel-1, which means the imagery with more input information to the deep learning models. Importantly, it is only Sentinel-1 that offers open operational data with up to every 6 days repeat. Because of this, the discussed approaches developed and tested specifically for PolSAR imagery at a higher resolution cannot be considered applicable for a wide-area mapping, yet. Similarly, Ahishali \textit{et al.} \cite{ahishali2019dual} applied end-to-end approaches to SAR data. They have also worked with single polarized COSMO-SkyMed imagery. However, all the imagery they considered was X-band SAR contrary to C-band imagery we use here and again only on a small scale. The authors proposed a compact CNN model that they found had outperformed some of the off-the-shelf CNN methods, such as Xception and Inception-ResNet-v2. It is important to note that compared to those, the off-the-shelf models that we consider here are more sophisticated semantic segmentation models, some which employ Xception or ResNet but only as a module in their feature extraction parts.

In summary, the capabilities of the deep learning approaches for the classification have been investigated to a lesser extent for SAR imagery than for optical imagery. The attempts to use SAR data for land cover classification were relatively limited in scope, area, or the number of used SAR scenes. Particularly, wide-area land cover mapping was never addressed.
The reasons for this include comparatively poor availability of SAR data compared to optical (greatly changed since the advent of Sentinel-1), complex scattering mechanisms leading to ambiguous SAR signatures for different classes (which makes SAR image segmentation more difficult than the optical image segmentation \cite{li2015homogeneous}), as well as the speckle noise caused by the coherent nature of the SAR imaging process. 
%%%%%%%%%%%%%%%%%%%%%%%%%%%%%%%%%%%%%%%%%%%%%%%%%%%%%%%%%%%%%%%%%%%%%%%%%%%%%%%%%%%%%%%%%%%%%%%%%%%%%%%%%%%%%%%%%%%%%%%%%%%%%%
\subsection{Study goals}
Present study addresses the identified research gap of a lack of wide-area land cover mapping using SAR data. We achieve this by training, fine-tuning, and evaluating a set of suitable state-of-the-art deep learning models from the class of semantic segmentation models, and demonstrating their \textit{suitability for land cover mapping}. Moreover, our work is the first to examine and demonstrate the suitability of deep learning for land cover mapping from SAR images on a \textit{large-scale, i.e., across the whole country}. 

Specifically, we applied the semantic segmentation models on the SAR images taken over Finland. We focused on the images of Finland because there is the land cover mask of a suitable resolution that can be used for training labels (i.e., CORINE). The training is performed with the seven selected models (SegNet \cite{badrinarayanan2017segnet}, PSPNet \cite{zhao2017pyramid}, BiSeNet \cite{yu2018bisenet}, DeepLabV3+ \cite{chen2018encoder,chen2018deeplab}, U-Net \cite{RFB15a,howard2017mobilenets}, FRRN-B \cite{pohlen2017full}, and FC-DenseNet \cite{jegou2017one}), which have encoder modules \textit{pre-trained} on the large RGB image corpus ImageNet 2012.\footnote{http://image-net.org/challenges/LSVRC/2012}  Those models are freely available.\footnote{https://github.com/tensorflow/models/tree/master/research/slim\#pre-trained-models} In other words, we reused semantic segmentation architectures developed for natural images with pre-trained weights on RGB images and we fine-tuned them on the SAR images.
%, hence applying \textit{transfer learning} from regular RGB images to the SAR images. 
Our results (with over 90\% overall accuracy) demonstrate the effectiveness of the deep learning methods for the land cover mapping with SAR data.

In addition to having the \textit{high-resolution CORINE map that can serve as a ground-truth (labels)} for training the deep learning models, another reason that we selected Finland is that it is a \textit{northern country with frequent cloud cover}, which means that using optical imagery for wide-area mapping is often not feasible. Hence, demonstrating the usability of radar imagery for land cover mapping is particularly useful here. 

Even though Finland is a relatively small country, there is still considerable heterogeneity present in terms of  land cover types and how they appear in the SAR images. Namely, \textit{SAR backscattering is sensitive to several factors that likely differ between countries or between distant areas within a country}. Examples of such factors are moisture levels, terrain variation and soil roughness, predominant forest biome and tree species proportions, types of shorter vegetation and crops in agricultural areas, and specific types of built environments. We did not contain our study to a particular area of Finland where the SAR signatures might be consistent but we obtained the images across a wide area. Hence, demonstrating the suitability of our methods in this setting hints at their potential generalizability. Namely, it means that, similarly as we did here, the semantic segmentation models can be fine-tuned and adapted to work on data from other regions or countries with the different SAR signatures.

On the other hand, we took into account that the same areas \textit{will appear somewhat different on the SAR images across different seasons}. Scattering characteristics of many land cover classes change considerably between the summer and winter months, and sometimes even within weeks during seasonal changes \cite{antropov2012, antropov2014}. These include snow cover and melting, freeze/thaw of soils, ice on rivers and lakes, crops growing cycle, leaf-on and leaf-off conditions in deciduous trees. Because of this, in the present study, we focused only on the scenes acquired during the summer season. However, we did allow our training dataset to contain several images of the same area, taken during different times during the summer season. This way not only spatial, but also temporal variation of SAR signatures is introduced.

Our contributions can be summarised as follows:
\begin{description}
	\item[C1:] We thoroughly \textit{benchmarked} seven selected state-of-the-art semantic segmentation models covering a diverse set of approaches for land cover mapping using Sentinel-1 SAR imagery. We provided insights on the best models in terms of both \textit{accuracy} and \textit{efficiency}.
	\item[C2:] Our results demonstrated the power of deep learning models along with SAR imagery for accurate \textit{wide-area land cover mapping} in the \textit{cloud obscured boreal zone and polar regions} These results can serve as \emph{baselines} when developing new, specialized approaches to SAR imagery.
\end{description}

\section{Deep Learning Terminology}
%%%%%%%%%%%%%%%%%%%%%%%%%%%%%%%%%%%%%%%%%%%%%%%%%%%%%%%%%%%%%%%%%%%%%%%%%%%%%%%%%%%%%%%%%%%%
As with other \textit{representation learning} models, the power of deep learning models comes from their ability to  \textit{learn rich features} (representations) from the dataset automatically \cite{goodfellow2016deep}. 
The automatically learned features are usually better suited for the classifier or other task at hand than hand-engineered features. Moreover, thanks to a \textit{large number of layers} employed, it has been proven that the deep learning networks can discover \textit{hierarchical representations}, so that the higher level representations are expressed in terms of the lower level, simpler ones. For example, in the case of images, the low-level representations that can be discovered are edges, and using them, the mid-level ones can be expressed, such as corners and shapes, and this helps to express the high-level representations, such as object elements and their identities \cite{goodfellow2016deep}.

The deep learning models in computer vision can be grouped according to their main task in three categories. In Table \ref{t:term}, we provide a description for those categories. However, the deep learning terminology for those tasks does not always correspond well to the terminology used in the remote sensing community. Relevant to our task, a number of remote sensing studies uses the term \textit{classification} in the context of land cover mapping, inherently meaning pixel- or region-based classification, which in the deep learning terminology corresponds to \textit{semantic segmentation}. In Table \ref{t:term} we list the corresponding terminology that we encountered being used for each task in both, the deep learning and remote sensing communities. This is helpful to disambiguate when talking about different, and recognize when talking about the same tasks in the two domains. In the present study, the focus is on land cover mapping. Hence, we tackle \textit{semantic segmentation} in the deep learning terminology and \textit{image classification, i.e., pixel-wise classification}, in the remote sensing terminology. \rev{Another terminology issue that often arises is about the dataset types used. The dataset that is held out from the training set and used to give an estimate of the model's performance during the training phase is referred to as a \emph{development dataset} or \emph{validation dataset} in the deep learning context. From remote sensing viewpoint, both training and development/validation datasets belong to training phase data. Further, the term \emph{validation data} in remote sensing context is typically reserved for the datasets used during the final evaluation (accuracy assessment) on completely independent data not involved in the training phase, i.e., what is called a \emph{test dataset} in deep learning. Hence, to avoid any confusion, we will avoid using the \emph{validation} term in the text, calling respective datasets as training, development, and test (accuracy assessment) data.}

%\textcolor{red}{SANJA or OLEG: here the validation/development vs accuracy assessment /testing paragraph to be added.}

{ % begin box to localize effect of arraystretch change
	\renewcommand{\arraystretch}{2.0}
	\begin{table}[]\caption{Terminology for the main tasks in computer vision and its use in the deep learning versus remote sensing communities.\label{t:term}}
		\begin{tabular}{p{0.17\linewidth}|p{0.22\linewidth}|p{0.44\linewidth}}
			 \textbf{Deep learning} & \textbf{Remote sensing}  & \textbf{Task description}  \\ \hline
			Classification \cite{krizhevsky2012imagenet} & Image Annotation, \break Scene Understanding, Scene Classification & Assigning a \textit{whole image} to a class based on what is (mainly) represented in it, for example a ship, oil tank, sea or land. \\
			Object Detection, Localization, Recognition \cite{goodfellow2016deep} & Automatic Target \break Recognition & Detecting (and localizing) presence of particular \textit{objects} in an image. These algorithms can detect several objects in the given image. For instance ship detection in SAR images.\\
		Semantic Segmentation \cite{long2015fully} & Image Classification, \break Clustering & Assigning a class to each \textit{pixel} in an image based on which image object or region it belongs to. These algorithms not only detect and localize objects in the image, but also output their exact areas and boundaries. \\
		\end{tabular}
	\end{table}
}

\textit{Convolutional Neural Networks (CNNs)} \cite{lecun1998gradient,krizhevsky2012imagenet} are the deep learning models that have transformed the computer vision field. Initially, CNNs are defined to tackle the \textit{image classification} (deep learning terminology) task. Their structure is inspired by the visual perception of mammals \cite{hubel1962receptive}. CNNs are named after one of the most important operations, which is particular to them compared to other neural networks, i.e., \textit{convolutions}. Mathematically, a convolution is a combination of two other functions. A convolution is applied on the image by sliding a \textit{filter (kernel)} of a given size $k \times k$ which is usually small compared to the original image size. Different purpose filters are designed; for example, a filter can serve as a vertical edge detector. Application of such a convolution operation on an image results in a feature map. Another common operation that is usually applied after a convolution is \textit{pooling}. Pooling reduces the size of the feature map while providing robustness to the extracted features. Common CNNs end with a \textit{fully connected layer} which is used for final predictions, commonly for image classification.  By employing a large number of convolutional layers (depth), CNNs are able to extract gradually more complex and abstract features. The first CNN model to demonstrate \rev{its} impressive effectiveness in image classification (of hand digits) was LeNet \cite{lecun1998gradient}. Several years later, Krizhevsky \textit{et al.} \cite{krizhevsky2012imagenet} developed AlexNet, the deep CNN to dramatically push the limits of classification accuracy on the famous ImageNet computer vision challenge \cite{russakovsky2015imagenet}.
Since then, a variety of CNN-based models are proposed. Some notable examples are: VGG network \cite{Simonyan2014VeryDC}, ResNet \cite{he2016deep}, DenseNet \cite{huang2017densely}, and Inception V3 \cite{szegedy2015going}. The effectiveness of CNNs has been also proven in various real-world applications \cite{ji20133d,sainath2013deep}.

\begin{figure}[!htb]
  	\centering
	\includegraphics[width=0.9\linewidth]{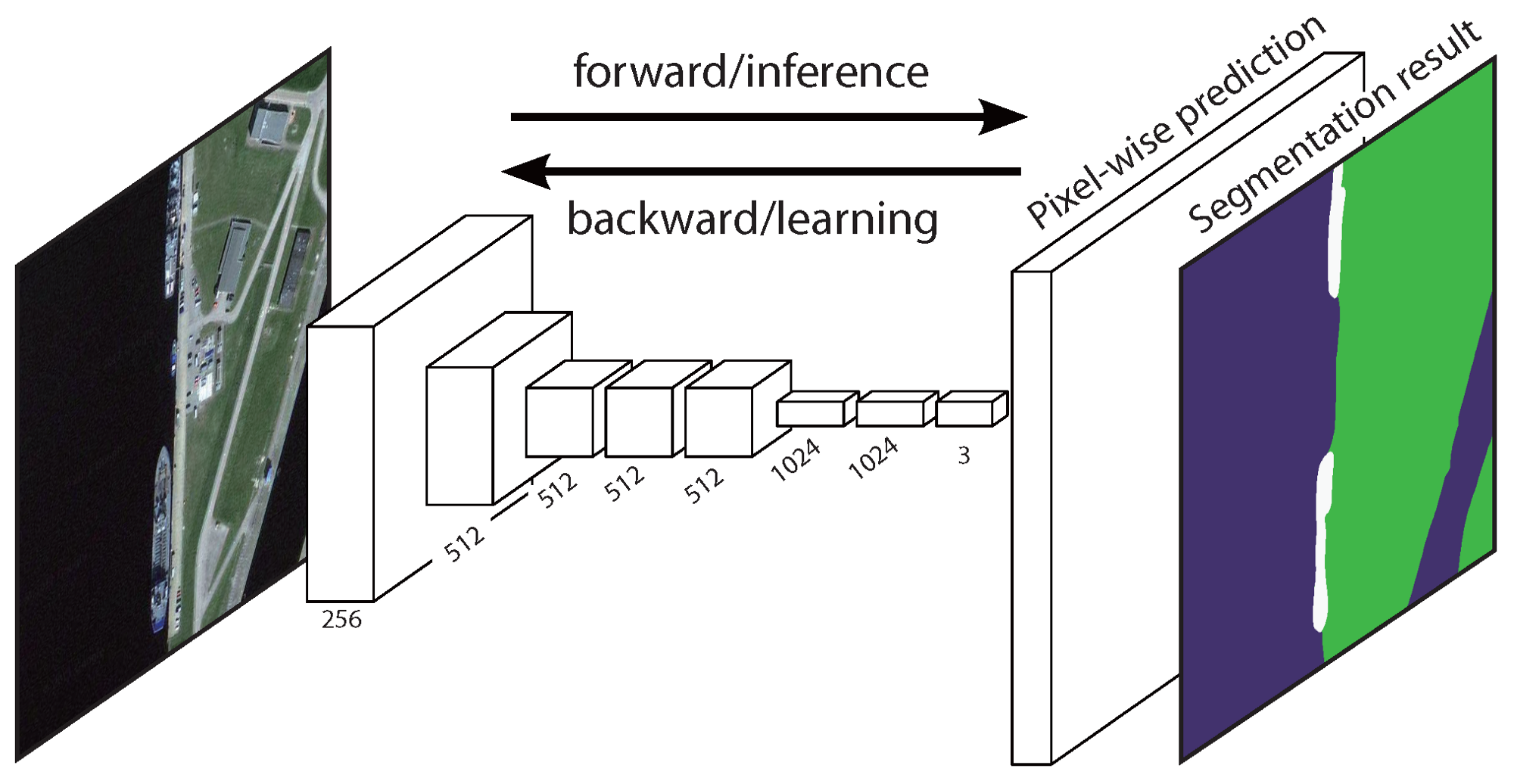}%
	\caption{The architecture of Fully Convolutional Neural Networks (FCNs) \protect{\cite{long2015fully}}\label{fig:FCNN}}
\end{figure}

Once CNNs have proven their effectiveness to classify images, Long \textit{et al.} \cite{long2015fully} were the first to discover how they can augment a given CNN model to make it suitable for the \textit{semantic segmentation} task -- they proposed the \textit{Fully Convolutional Neural Network (FCN)} framework. 
 This generic architecture can be used to adapt any CNN network used for classification into a segmentation model. 
Namely, the authors have shown that by replacing the last, fully connected layer, with an appropriate convolutions layer, so that they will upsample and restore the resolution of the input at the output layer, CNNs can be transformed to classify each individual pixel (instead of the whole image). 
The basic idea is as follows. The \textit{encoder} is used to learn the feature maps, and is usually based on a pre-trained deep CNN for classification, such as ResNet, VGG, or Inception. The \textit{decoder} part serves to upsample the discriminative features that the encoder has learned from the coarse-level feature map to the fine, pixel level. Long \textit{et al.} \cite{long2015fully} have shown that this upsampling (backward) computation can be efficiently performed using backward convolutions (deconvolutions). Moreover, this means that the specific CNN models, such as those mentioned above, can all be incorporated in the FCN framework for segmentation, giving rise to FCN-AlexNet \cite{long2015fully}, FCN-ResNet \cite{he2016deep}, FCN-VGG16 \cite{long2015fully}, FCN-DenseNet \cite{jegou2017one} etc.  We present a diagram of the generic FCN architecture in Figure \ref{fig:FCNN}. 

\section{Materials and methods}
\label{sec:data}
%%%%%%%%%%%%%%%%%%%%%%%%%%%%%%%%%%%%%%%%%%%%%%%%%%%%%%%%%%%%%%'
Here, we first describe the study site, SAR, and reference data. This is followed by an in-depth description of the deep learning terminology and the models used in the study. We finish with the description of the experimental setup and the evaluation metrics.

%%%%%%%%%%%%%%%%%%%%%%%%%%%%%%%%%%%%%%%%%%%%%%%%%%%%%%%%%%%%%%

\subsection{Study site}

\begin{figure*}
		\centering
		\includegraphics[width=0.9\linewidth]{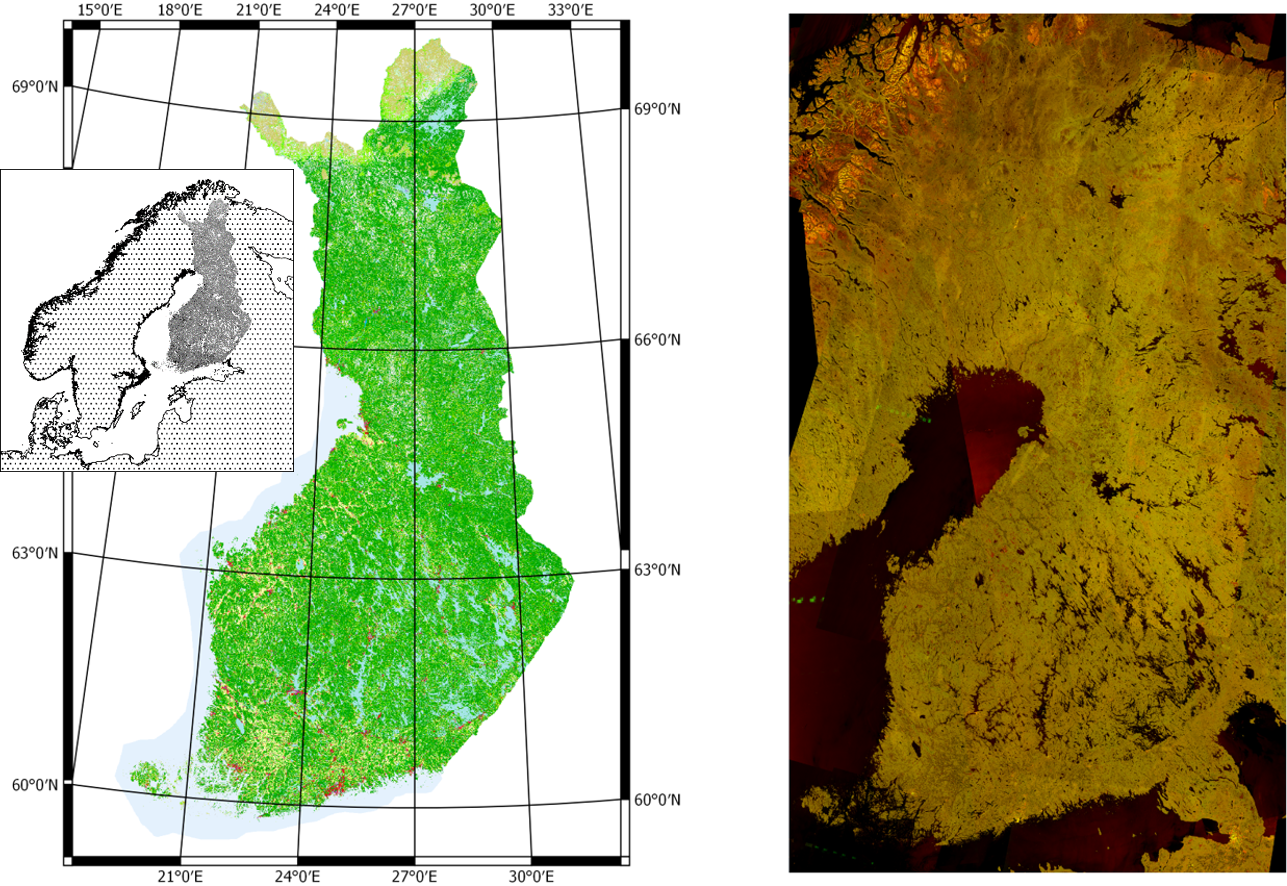}
	\caption{Study area in Finland: (a) reference CORINE land cover data; (b) example of compiled Sentinel-1 SAR mosaic that includes the whole country.}
	\label{fig:study_area_ref_data}
\end{figure*}

\rev{Our study site is covering the territory of Finland located to the south of 66.0\degree  latitude, that is effectively whole country without Finnish Lapland. The study area is shown in Figure \ref{fig:study_area_ref_data}. Southern Finland is primarily covered by boreal forests with lakes, marshes, open bogs, agricultural areas and urban settlements. We have omitted Lapland due to considerably different land cover composition and topography compared to the rest of the country. The terrain height variation within the study area is moderate and mostly within $100-300$ meters range.}

\subsection{SAR data}
Presently, Sentinel-1 is a C-band SAR dual-satellite system with two satellites orbiting $180^{\circ}$ apart \cite{torres2012}, launched in 2014 and 2016, respectively. The operational acquisition modes are Stripmap (SM), Interferometric Wide-Swath (IW), Extra Wide Swath (EW), and Wave Mode (WV). The IW-mode is the default mode over land, providing 250 km wide swath composed of three sub-swaths, with single look image at 5 m by 20 m spatial resolution. It uses the so-called TOPS (Terrain Observation with Progressive Scan) SAR mode.  

{The SAR data acquired by Sentinel-1 satellites in IW mode are used in our study. Specifically, we used only Sentinel-1A imagery acquired during the summer 2018.}

{Original scenes were downloaded as Level-1 Ground Range Detected (GRD) products. They represent focused SAR data that has been detected, multi-looked and projected to ground-range using an Earth ellipsoid. The images were orthorectified using the Technical Research Centre of Finland (VTT) in-house software employing the local digital terrain model (with 20 m resolution) available from National Land Survey of Finland. The pixel spacing of orthorectified scenes was set to 20 m. Ortho-rectification included terrain flattening to obtain the backscatter signal in gamma-nought format \cite{small2012terrain}. The scenes were further re-projected to the ERTS89 / ETRS-TM35FIN projection (EPSG:3067) and resampled to a final pixel size of 20 metres.}

{The Sentinel-1 images were mosaiced into 7 homogeneous SAR mosaics covering whole Finland. Each mosaic was compiled from approximately 90 Sentinel-1 IW scenes (both ascending and descending paths),  and it takes about 12 days to collect enough imagery to have the whole country covered. Altogether seven SAR mosaics were produced during the summer 2018. These SAR mosaics are further used for sampling the training, development, and testing images that are input to Deep Learning models as described in detail in Section \ref{exp-set}.}

{The geographical coverage of each SAR mosaic is shown in Figure \ref{fig:study_area_ref_data}.}

\subsection{Reference data}
\begin{figure*}
  \includegraphics[width=\linewidth]{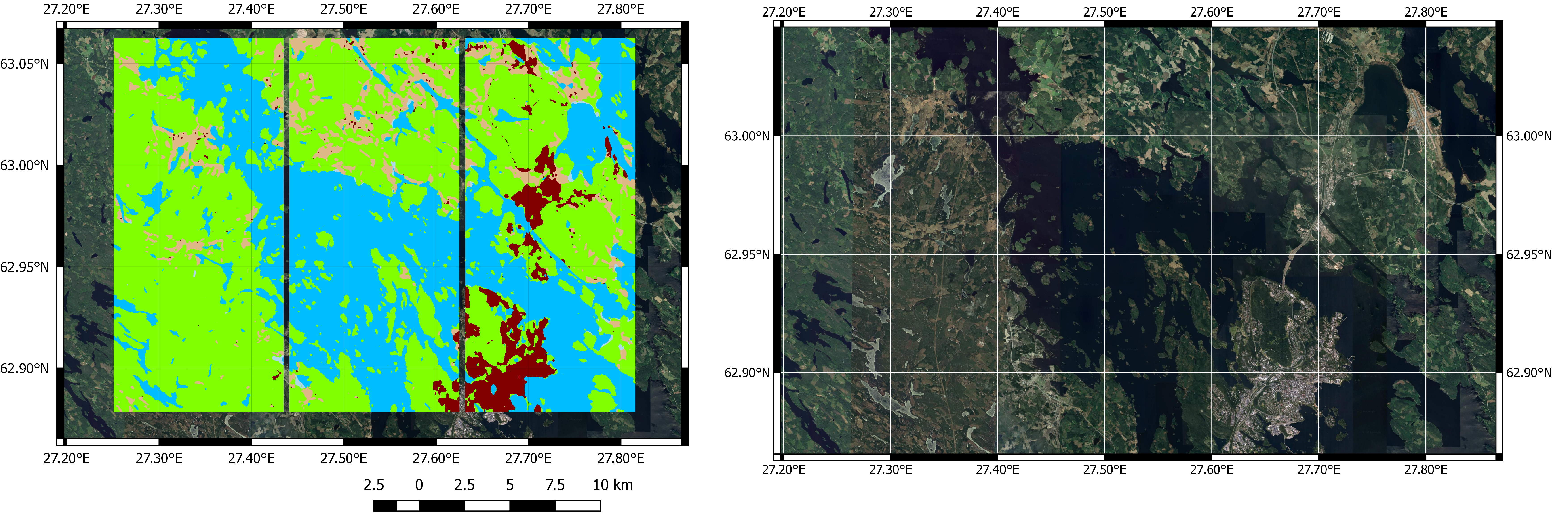}
\caption{Zoomed in area fragment with our reference data, i.e., CORINE shown on top (left) along with the Google Earth layer (right).}
\label{fig:zoom_in_study_area_ref_data}
\end{figure*}

 In Finland, the Finnish Environment Institute (SYKE) is responsible for production of the CORINE maps. While for most of the EU territory, the CORINE mask of $100\ m\times100\ m$ spatial resolution is available, the national institutions might choose to create more precise maps, and SYKE, in particular, had produced a $20\ m\times20\ m$ spatial resolution mask for Finland (Figure \ref{fig:zoom_in_study_area_ref_data}), { with the first one in 2000.  Since then, the updates have been produced regularly, with the latest one CLC2018 that well corresponds to the acquisition timing of our SAR data. There are $48$ different land use classes in the map that can be hierarchically grouped into 4 CLC Levels. In detail, there are $30$ classes on CLC Level-3, $15$ classes on CLC Level-2, and $5$ top CLC Level-1 classes. According to the information provided by SYKE for CLC2012, the accuracy of the CLC Level-3 was $61\%$, of the CLC Level-2, $83\%$, and of the CLC Level-1, it was $93\%$. In this study, we use its updated and revised version, CLC2018, having good results on both internal and external quality control.\footnote{\url{https://www.syke.fi/en-US/Research__Development/Research_and_development_projects/Projects/Producing_land_cover_and_land_use_data_in_CORINE_Land_Cover_2018_project_in_Finland}} The selected classes and their corresponding color codes used for our segmentation results are shown in Table \ref{t:corine_classes}. Our superclasses generally correspond to CLC Level-1 classes, with minor corrections for ``artificial surfaces" class that is not fully included in urban class, but some elements are distributed to other classes; most notably green urban areas were included to forest class in our study as those are essentially parks and and mixed boreal forestland enclosed within urban-designated areas. }

Until the most recent CORINE production round, EEA member countries adopted national approaches for the production of CORINE. EEA Technical Guidelines include manual digitalization of land cover change based on visual interpretation of optical satellite imagery.  In Finland, the European CLC was not applicable for the majority of national users due to large minimal mapping unit (MMU). 
Thus national version was produced with somewhat modified nomenclature of classes \cite{finalndclc1,finalndclc2}. The national high-resolution CLC2018 data is in raster format of 20 m, with corresponding MMU.
In the provision of 2018 update of CLC, obtaining optical imagery over Scandinavia and Britain was particularly challenging because of the frequent clouds, thus \textit{calling for the use of radar imagery} to meet user requirements on accuracy and coverage \cite{balzter2015}. CORINE map itself is normally built from high resolution satellite images acquired primarily during the summer and, to a smaller extent, during the spring months \cite{buttner2004corine}.

%\begin{table}[]\caption{CORINE based land cover classes and their color codes used in our classification results\label{t:corine_classes}}
%	\centering
%\begin{tabular}{llllll}
%\textbf{land cover class} & \textbf{R} & \textbf{G} & \textbf{B} & \textbf{color}& \textbf{description} \\
%\hline
%Urban  	 & 128 & 0 & 0 & red& continuous urban fabric, industrial and commercial %units, construction sites \\
%Agriculture   & 222 & 184 & 135 & brown& \oleg{add description}\\
%Forested areas   & 127 & 255 & 0 & green& \oleg{add description}\\
%Peatland  & 173 & 216 & 230 & light blue& \oleg{add description}\\
%Water bodies  & 0 & 191 & 255 & blue& \oleg{add description}\\

%\end{tabular}
%\end{table}

\begin{table}[]
\caption{Description of CORINE based land cover classes and their map color codes \label{t:corine_classes}}
\begin{tabular}{m{1.7cm} m{1.2cm} m{4.6cm}}
LC class     & RGB         & description                                    \\ \hline
urban          & 128,0,0   & urban fabric, industrial and commercial units, constructions sites, dump sites \\
agriculture  & 222,184,135 & agricultural and agro-forestry areas, fruit trees and berry plantations, pastures \\
forested areas & 127,255,0 & \begin{tabular}[c]{@{}l@{}}broad-leaved, coniferous and mixed forest, \\ transitional woodland/shrub\end{tabular}         \\
peatland     & 173,216,230 & peatland, bogs, inland marshes and salt marshes  \\         
water bodies & 0,191,255   & rivers, lakes, sea                            
\end{tabular}
\end{table}

\subsection{Semantic Segmentation Models}
%%%%%%%%%%%%%%%%%%%%%%%%%%%%%%%%%%%%%%%%%%%%%%%%%%%%%%%%%%%%%%%%%%%%%%%%%%%%%%%%%%%%%%%%%%%%
We selected following seven state-of-the-art \cite{garcia2017review} semantic segmentation  models to test for our land cover mapping task: SegNet \cite{badrinarayanan2017segnet}, PSPNet \cite{zhao2017pyramid}, BiSeNet \cite{yu2018bisenet}, DeepLabV3+ \cite{chen2018encoder,chen2018deeplab}, U-Net \cite{RFB15a,howard2017mobilenets}, FRRN-B \cite{pohlen2017full}, and FC-DenseNet \cite{jegou2017one}. The models were selected to cover a wide set of approaches to semantic segmentation. In the following, we describe its specific architecture for each of these DL models. We will use the following common abbreviations: \texttt{conv} for convolution operation, \texttt{concat} for concatenation, \texttt{max pool} for max pooling operation,  \texttt{BN} for batch normalisation, and \texttt{ReLU} for the rectified linear unit activation function.
%%%%%%%%%%%%%%%%%%%%%%%%%

\subsubsection{ \textbf{ BiSeNet (Bilateral  Segmentation  Network)}}
  \begin{figure}[!htb]{}
  	\centering
	\includegraphics[width=0.7\linewidth]{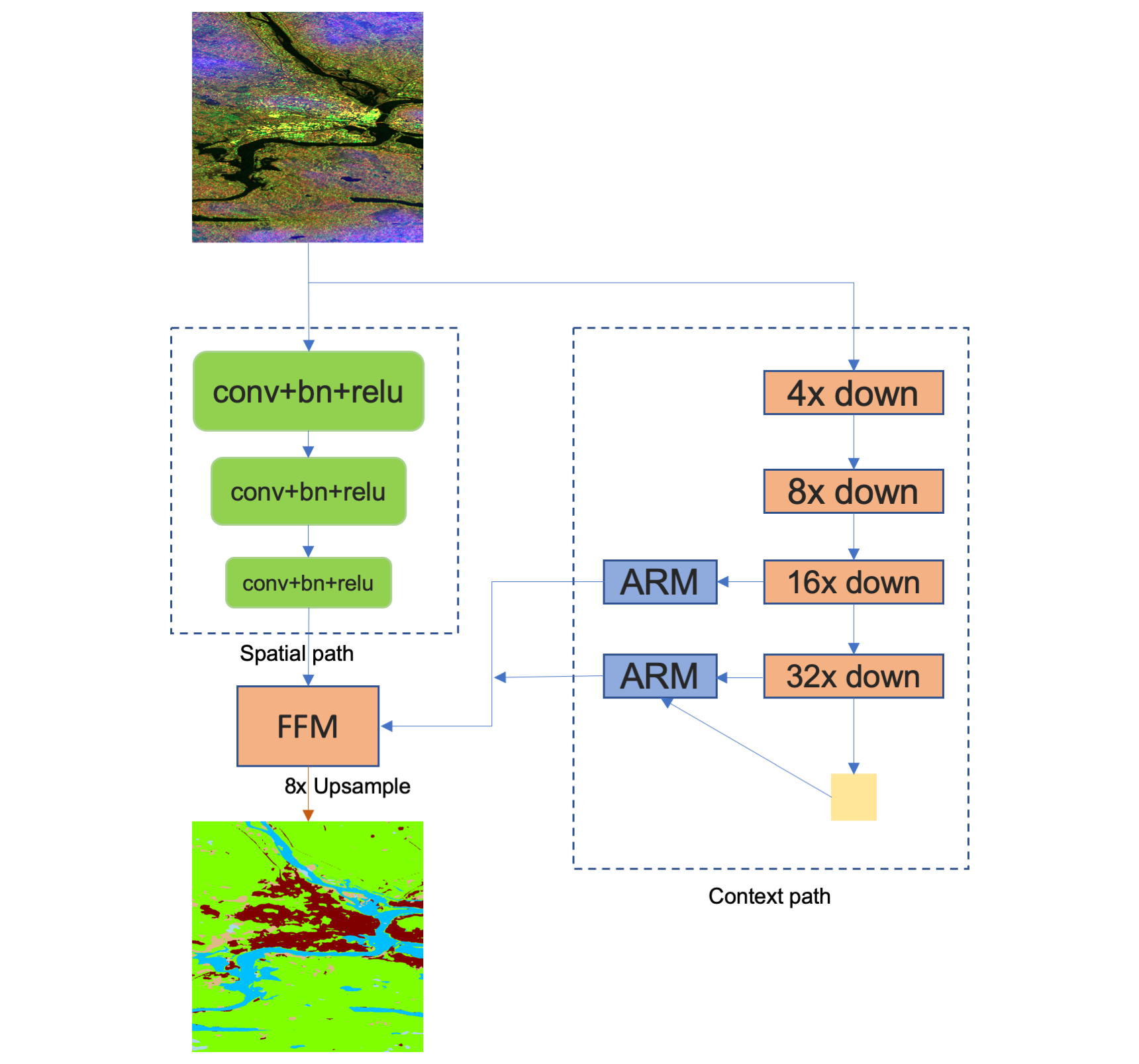}%
	\caption{The architecture of BiSeNet. ARM stands for the Attention Refinement Module and FFM for the Feature Fusion Module introduced in the model's paper \protect{\cite{yu2018bisenet}}. \label{fig:bisenet}}%
\end{figure}

BiSeNet model is designed to decouple the functions of 
encoding additional spatial information and enlarging the
receptive field, which are fundamental to achieving good segmentation performance.
As can be seen in Figure \ref{fig:bisenet}, there are two main components to this model:  Spatial Path (SP) and Context Path (CP).
Spatial Path serves to encode rich spatial information.
Context Path serves to provide sufficient receptive field and uses global average pooling and pre-trained Xception \cite{chollet2017xception} or ResNet \cite{he2016deep} as the backbone. 
The goal of the creators was not only to obtain superior performance but to achieve a balance between the speed and performance. Hence, BiSeNet is a relatively fast semantic segmentation model.

\subsubsection{\textbf{SegNet (Encoder-Decoder-Skip)}}
  \begin{figure}[!htb]{}
  	\centering
	\includegraphics[width=\linewidth]{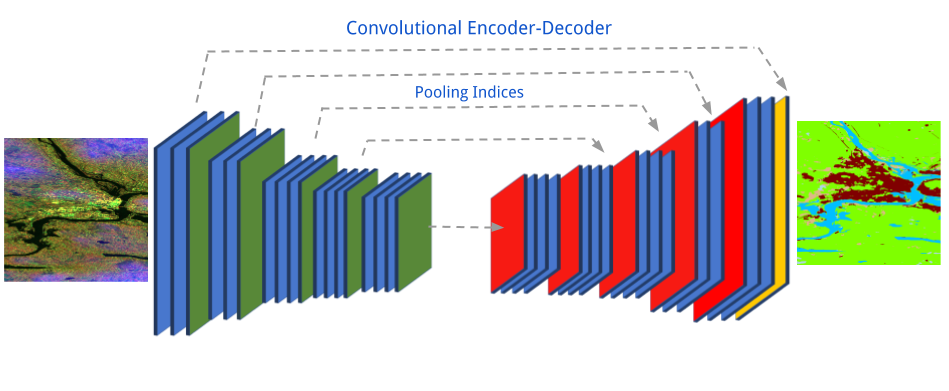}%
	\caption{The architecture of SegNet-based Encoder-Decoder with Skip connections \protect{\cite{badrinarayanan2017segnet}. Blue tiles represent Convolution + Batch Normalisation + ReLU, green tiles represent Pooling, red -- Upsampling, and yellow -- a softmax operation. }\label{fig:segnet}}%
\end{figure}
Similarly to BiSeNet, SegNet is also designed with computational performance in mind, this time, particularly during inference. Because of this, the network has a significantly smaller number of trainable parameters compared to most of the other architectures. The encoder in SegNet is based on VGG16: it consists of its first 13 convolutional layers, while the fully connected layers are omitted. Hence, the novelty of this network lies in its decoder part, as follows. The decoder consists of one decoder layer for each encoder layer and so it also has 13 layers. 
Each individual decoder layer utilizes max-pooling indices memorized from its corresponding encoder feature map. The \rev{authors have shown} that this enhances boundary delineation between classes.
Finally, the decoder output is sent to a multi-class
soft-max function yielding classification for each pixel (see Figure \ref{fig:segnet}). 

\subsubsection{\textbf{Mobile U-Net}}
\begin{figure}[!htb]{}
  	\centering
	\includegraphics[width=0.9\linewidth]{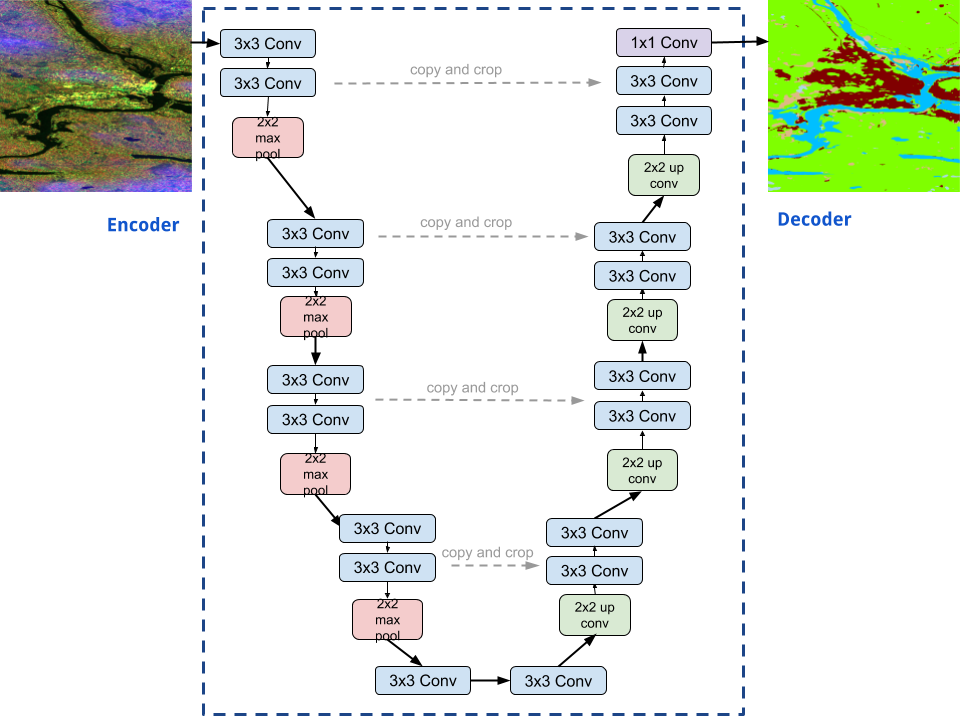}%
	\caption{The architecture of U-Net \protect{\cite{ronneberger2015u}}\label{fig:UNet}}
\end{figure}

Mobile U-Net is based on the U-Net \cite{ronneberger2015u} semantic segmentation architecture shown in Figure \ref{fig:UNet}. In designing U-Net, Fully Convolutional approach was generally employed with a following modification. Their upsampling part of the architecture has no fully convolutional layer but is nearly symmetrical to the feature extraction part due to the use of the similar feature maps. This results in a \textit{u-shaped} architecture (see Figure \ref{fig:UNet}), and hence the name of the model. While originally developed for biomedical images, the U-net architecture has proven successful for image segmentation in other domains, as well. Here, we somewhat modify the U-Net architecture, according to MobileNets \cite{howard2017mobilenets} framework, to improve its efficiency. In particular, the MobileNets framework uses Depthwise Separable Convolutions, a form which factorizes standard convolutions (e.g., $3 \times 3$) into a depthwise convolution (applied separately to each input band) and a pointwise ($1 \times 1$) convolution to combine the outputs of depthwise convolution.

\subsubsection{\textbf{DeepLab-V3+}}
\begin{figure}[!htb]{}
  	\centering
	\includegraphics[width=0.9\linewidth]{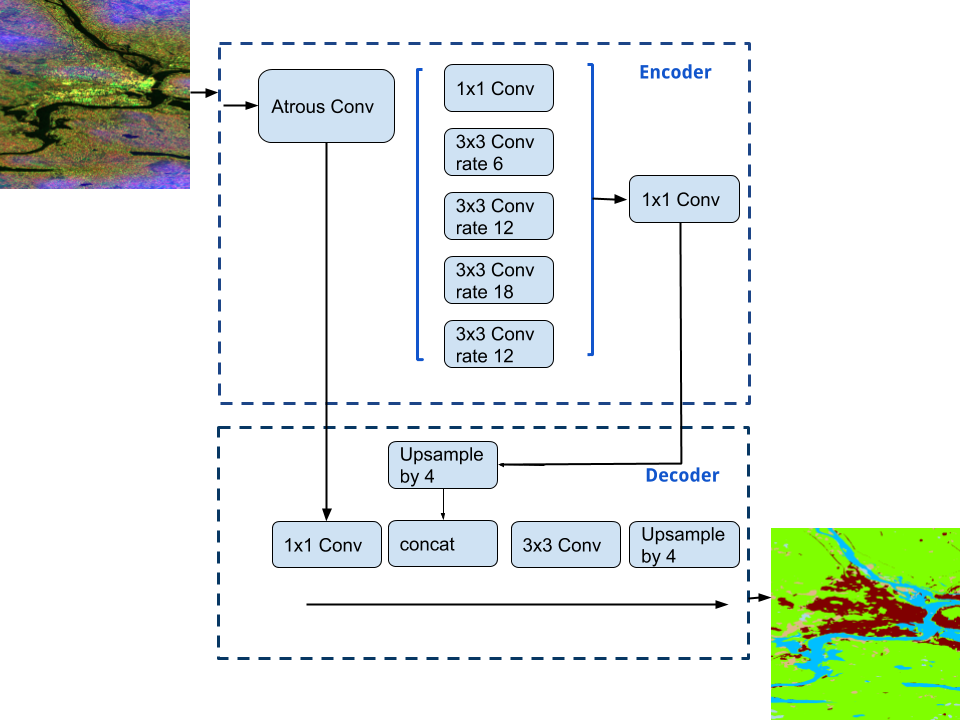}%
	\caption{The architecture of DeepLabV3+ \protect{\cite{chen2018encoder}}\label{fig:DeepLab}}%
\end{figure}

DeepLab-V3+ \cite{chen2018encoder} is an improved version of DeepLab-V3 \cite{chen2017rethinking}, while the latter is an improved version the original DeepLab \cite{chen2018deeplab} model. This segmentation model does not follow the FCN framework like the previously discussed models. The main features that distinguish the DeepLab model from FCNs are the \textit{atrous convolutions} for upsampling and the application of probabilistic machine learning models, concretely, \textit{conditional random fields (CRFs)} for a finer localization accuracy in the final fully connected layer. Atrous convolutions, in particular, allow to enlarge the context from which the next layer feature maps are learned, while preserving the number of parameters (and, thus, the same efficiency). Using a chain of atrous convolutions allows to compute the final output layer of a CNN at an arbitrarily high resolution (removing the need for the upsampling part as used in FCNs). In the follow up work, proposing DeepLab-V3, Chen \textit{et al.} \cite{chen2017rethinking} change the approach to atrous convolutions to gradually double the atrous rates, and show that with an adapted version, their new algorithm outperforms the previous one, even without including the fully connected CRF layer. Finally, in their newest adaption to the model, called DeepLab-V3+, Chen \textit{et al.} \cite{chen2018encoder} turn to a similar approach to the FCNs, i.e., they add a decoder module to the architecture (see Figure \ref{fig:DeepLab}). That is, they employ the features extracted by the DeepLab-V3 module in the encoder part, and add the decoder module consisting of $1 \times 1$ and $3 \times 3$ convolutions.

 \begin{figure}[!htb]{}
  	\centering
	\includegraphics[width=0.44\linewidth]{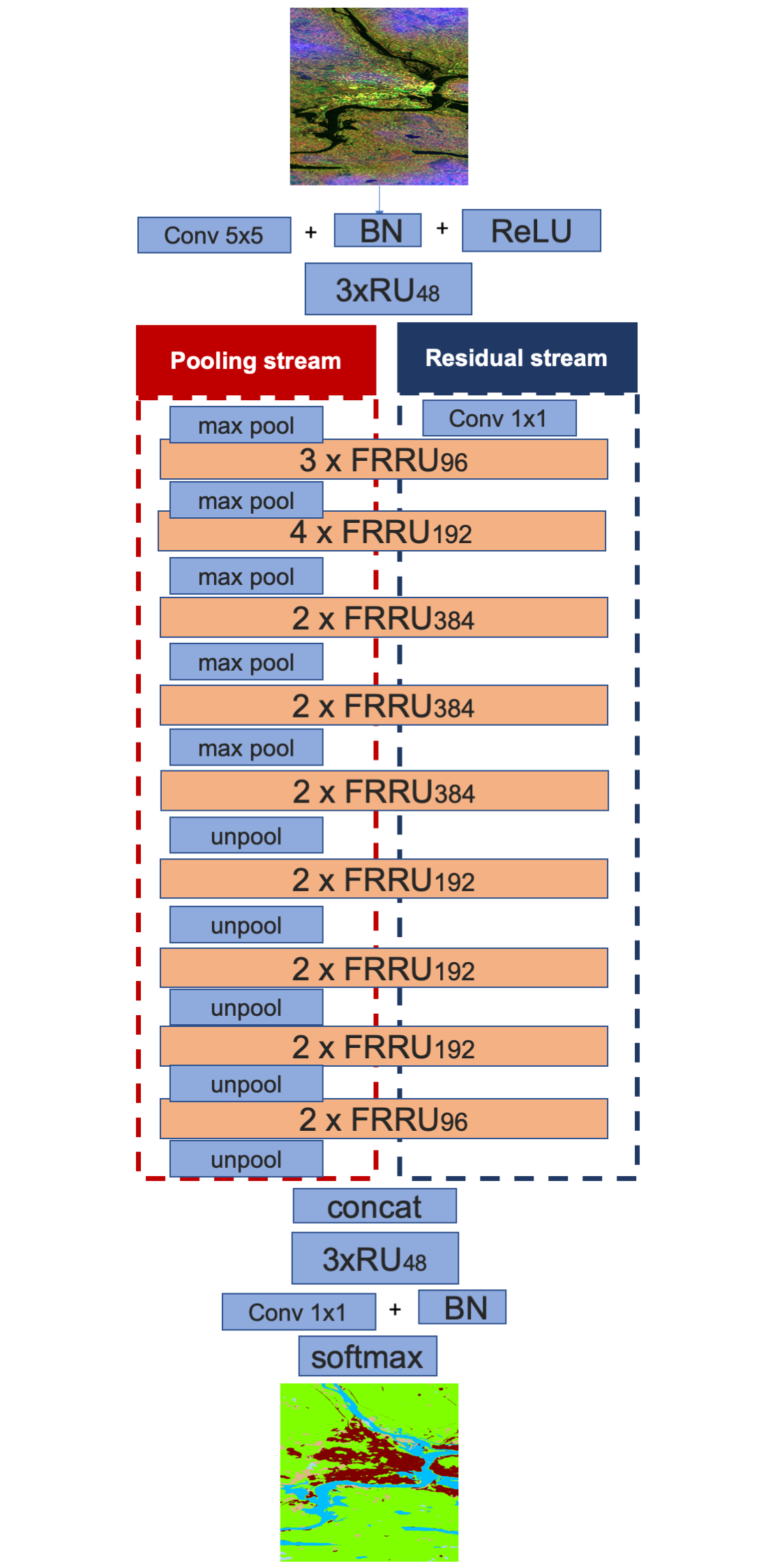}%
	\caption{The architecture of FRRN-B. RU\_n and FRRU\_n stand for residual units  and full-resolution residual units with n-channel convolutions, respectively. FRRUs simultaneously operate on the two streams \protect{\cite{pohlen2017full}}. \label{fig:frrnb}}%
\end{figure}

\subsubsection{\textbf{FRRN-B (Full-Resolution Residual Networks)}}
As we have seen, most of the semantic segmentation architectures are based on some form a FCN, and so they utilize existing classification networks, such on ResNet or VGG16 as encoders. We also discussed the main reason for such approaches, which is to take advantage of the learned weights from those architectures pretrained for the classification task. Nevertheless, one disadvantage of the FCN approach is that the resulting network outputs of the encoder part (particularly, after the pooling operations) are at a lower resolution, which deteriorates localization performance of the overall segmentation model. Pohlen \textit{et al.} \cite{pohlen2017full} proposed to tackle this by having two parallel network streams processing the input image: a pooling and a residual stream (Figure \ref{fig:frrnb}). As the name says, the pooling stream performs successive pooling and then unpooling operations, and it serves to obtain good recognition of the objects and classes. The residual stream computes residuals at the full image
resolution, which enables that low level features, i.e., object pixel-level locations, are propagated to the network output. 
The name of the model comes from its building blocks, i.e., full-resolution residual units. Each such unit simultaneously operates on the pooling and the residual stream. In the original paper \cite{pohlen2017full}, the authors propose two alternative architectures FRRN-A and FRRN-B, and they show that FRRN-B achieves superior performance on the Cityscapes benchmark dataset. Hence, we employ the FRRN-B architecture.

\subsubsection{\textbf{PSPNet (Pyramid Scene Parsing Network)}}
  \begin{figure}[!htb]{}
  	\centering
	\includegraphics[width=\linewidth]{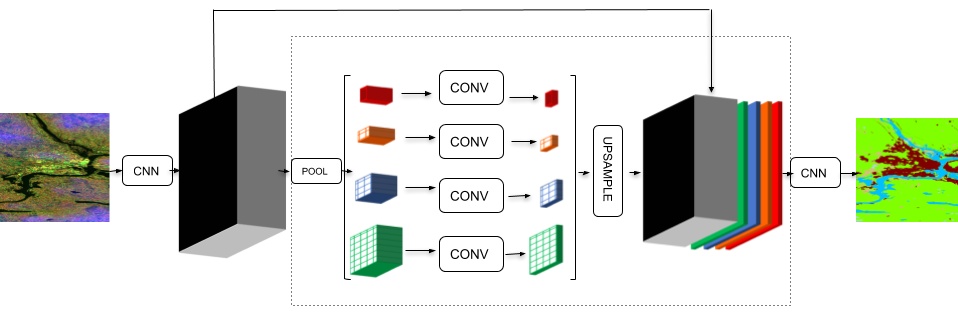}%
	\caption{The architecture of PSPNet \protect{\cite{zhao2017pyramid}}\label{fig:pspnet}}%
\end{figure}
Zhao \textit{et al.} \cite{zhao2017pyramid} propose the Pyramid Scene Parsing as a solution to the challenge of making the local predictions based on the local context only, and not considering the global image scene. In remote sensing, an example for this challenge happening could be when a model wrongly predicts the water with waves present in it as the dry vegetation class, because they appear similar and the model did not consider that these pixels are being part of a larger water surface, i.e., it missed the global context. In similarity to the other FCN-based approaches, PSPNet uses a pre-trained classification architecture to extract the feature map, in this case, ResNet. The main module of this network is the pyramid pooling, which is enclosed by a square in Figure \ref{fig:pspnet}. As can be seen in the Figure, this module fuses features at four scales: from the coarse (red) to the fine (green). Hence, the output of each level in the pyramid pooling module contains the feature map of a different resolution. In the end, the different features are stacked together yielding the final pyramid pooling global feature for predictions.

\subsubsection{\textbf{FC-DenseNet (Fully Convolutional DenseNets)}} 
\begin{figure}[!htb]
  	\centering
	\includegraphics[width=\linewidth]{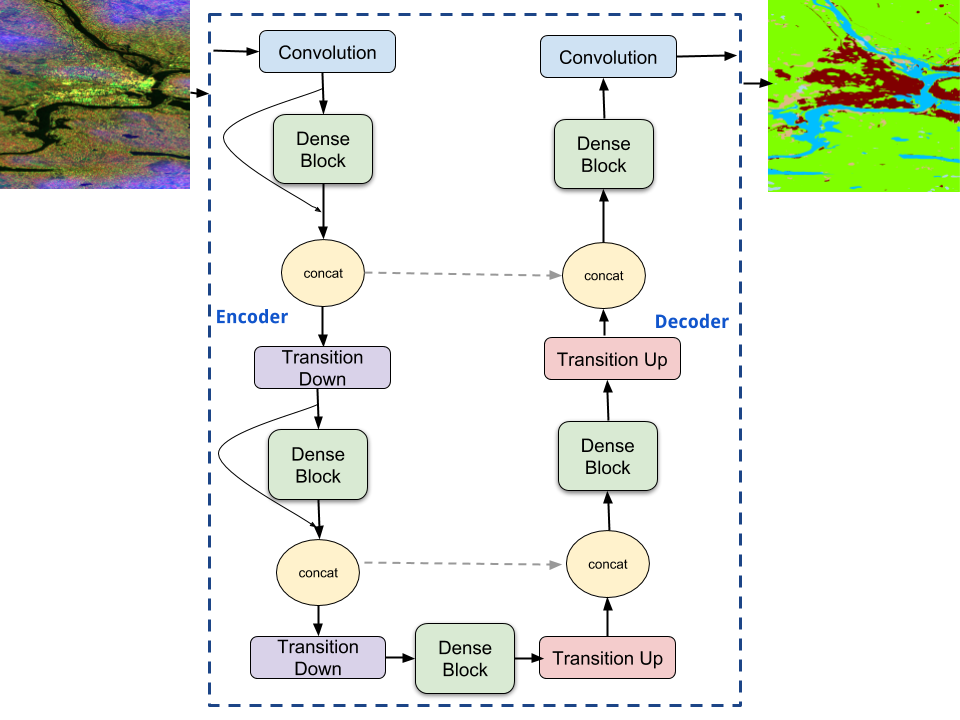}%
	\caption{The architecture of FC-DenseNet \protect{\cite{jegou2017one}}\label{fig:FC-DenseNet}}%
\end{figure}

This semantic segmentation algorithm is built using DenseNet CNN \cite{huang2017densely} as a basis for the encoder, followed by applying the FCN approach \cite{jegou2017one}.  The specificity of the DenseNet architecture is the presence of blocks where each layer is connected to all other layers in a feed-forward manner. Figure \ref{fig:FC-DenseNet} shows the architecture of FC-DenseNet where the blocks are represented by the \textit{Dense Block} units. According to \cite{huang2017densely}, such architecture scales well to hundreds of layers without any optimization issues, while yielding excellent results in classification tasks.
In order to efficiently upsample the DenseNet feature maps, Jegou \textit{et al.} \cite{jegou2017one} substitute the upsampling convolutions of FCNs by \textit{Dense Blocks} and \textit{Transitions Up}. The \textit{Transition Up} modules consist of transposed convolutions, which are then concatenated with the outputs from the input skip connection (the dashed lines in Figure \ref{fig:FC-DenseNet}).

\subsection{Training approach}
%The performance of supervised machine learning strongly depends on the availability of high-quality reference (labelled) data for training \cite{goodfellow2016deep,lecun2015deep}. This is also the case with our selected semantic segmentation models.

To accomplish better segmentation performance, there is an option to pre-train the semantic segmentation models (in particular, their encoder modules) using a larger set of available images of another type (such as natural images). Using the model pre-trained with natural images to continue training with the limited set of SAR images, the knowledge becomes effectively transferred from the natural to the SAR task \cite{bengio2012deep}. To accomplish such transfer, we used the models whose encoders were pre-trained for the ImageNet classification task and fine-tuned them using our SAR dataset (described next).

%In this work, we tested the approach of fine-tuning the deep learning architectures trained for the RGB ImageNet segmentation task. Differently to \cite{mahdianpari2018very}, we used SAR instead of optical imagery, and studied land cover mapping across a country instead of a small selected test site.

\subsection{{Experimental Setup}, \label{exp-set}}
In this section, we first describe how we prepared the SAR images for training with the deep learning models that are originally designed for natural images, and then we provide the details of our models' implementation and the hardware setup used.

\subsubsection{SAR Data Preprocessing}
Sentinel-1 imagery comes in two polarization channels (VH and VV), each of them being particularly informative about certain types of land cover. Hence,  using their combination is expected to yield better land cover mapping results than using any of them independently. \rev{Moreover, previous works suggested benefits of employing DEM in land cover mapping \cite{zhao2016detailed}, so we experimented also with topographic DEM from the National Land Survey.} {To assess the marginal utility of adding the DEM layer as compared to using solely SAR data, we prepared two training datasets: one with SAR data only, and another one with a DEM layer.} 
%We describe the two datasets in more detail next.

{The backscatter amplitude for both polarizations (VH and VV) represented first two channels for both datasets. As the third channel, after some preliminary tests, we decided to include a VH-to-VV amplitude ratio (also known as cross-pol ratio). This dataset was called \textit{\textbf{RGB SAR Ratio}}. For the second dataset, a DEM layer was used as the third channel, thus this dataset was called \textit{\textbf{RGB SAR DEM}}. In addition, for the deep learning models, each band should be normalized so that the distribution of the pixel values would resemble a Gaussian distribution centered at zero. This is done to yield a faster convergence during the training. Hence, each channel was normalized by channel-specific calibration factor using percentile contrast stretching, with no more than 1\% of pixel values clipped. }

%\textcolor{blue}{\textbf{RGB SAR Ratio Dataset.} The SAR backscatter amplitude for both polarizations (VH and VV) represented first two channels in this dataset. As the third channel, after some preliminary tests, we decided to include a VH-to-VV amplitude ratio (also known as cross-pol ratio). In addition, for the deep learning models, each band should be normalized so that the distribution of the pixel values would resemble a Gaussian distribution centered at zero. This is done to yield a faster convergence during the training. Hence, each channel was normalized by channel-specific calibration factor using percentile contrast stretching, with no more than 1\% of pixel values clipped.}

%\textcolor{blue}{\textbf{RGB SAR DEM Dataset.} We applied a similar procedure as for the previously described dataset. The only difference is that we now employed the DEM layer for the third (B) channel.
%} \sanja{Oleg, check if you want to comment more here on the normalisation of the DEM layer.}

\rev{The naming of the two datasets comes from the process used to create the images in them. Namely, VH-pol data of a Sentinel-1 image is assigned to R and VV-pol to G channel. For the third, B channel, in each of the datasets we used either cross-pol ratio of Sentinel-1 data or the DEM layer, respectively. Given that the semantic segmentation models expect RGB pixel values in the range (0,255), we scaled the normalized channel values for both datasets to this range.}

\subsubsection{Train/Development and Test (Accuracy Assessment) Dataset }  
The original images from the needed to be split into $512\ px\times512\ px$ partial images (further in the text called \textit{imagelets}) used for model training and testing.  Thus, each imagelet represented an area of roughly $10\times10\ km^2$. The first reason for this preprocessing has to do with the squared shape: some of the selected models required squared-shaped images. Some other of the models were flexible with the image shape and size but we wanted to make the setups for all the models the same so that their results are comparable. The second reason for this preprocessing has to do with computational capacity: with our hardware setup (described below), this was the largest image size that we could work with. 

%\textcolor{red}{NEEDS REWRITE: Upon splitting the \textit{SAR RGB-DEM} images, we discarded those imagelets that were completely outside the land mass area, as well as those for which we did not have a complete CORINE label (such as if they fell in part outside the Finnish borders). This resulted in more than $7K$ imagelets of size $512px\times512px$.} 

 {Given the geography of Finland, for representative training data, it is useful to include imagelets from the whole country (including the large cities) aside from the Finnish Lapland, where the land classes are distinctly different. On the other hand, some noticeable differences are found also in the gradient from east to west of the country. Hence, to achieve a representative training dataset, we selected all imagelets between the longitudes of 25\degree and 29\degree  for the accuracy assessment (so-called ``unobserved data" for model testing), and all the other imagelets we used for the model training (that is training/validation in the computer vision terminology). In this way, we prevented the situation in which two images of the same area but acquired at different times were used one for training and the other one for testing. In other words, we kept our training/development and test sets completely independent from each other.}
 
\begin{figure*}[!htb]
  	\centering
	\includegraphics[width=0.9\linewidth]{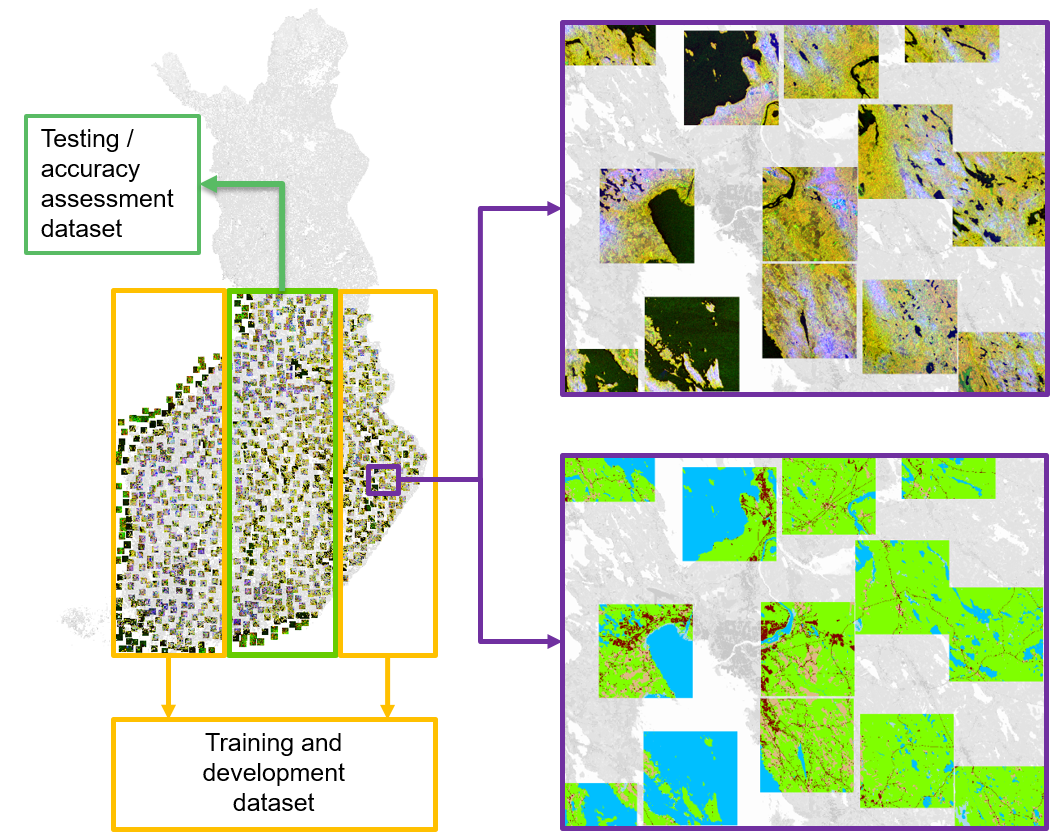}%
	\caption{\rev{The sampling of SAR and land cover imagelets and division into training \& development and testing datasets}}%
	\label{fig:sampling}
\end{figure*}

  {The areas for training/development and model testing are shown in Figure \ref{fig:sampling}. From each of the seven SAR mosaics, 1000 imagelets were generated using random sampling, while controlling for no spatial overlap between the imagelets. Among those 1000 imagelets, 400 were sampled from the testing area and set aside for the accuracy assessment, while the remaining 600 were sampled from the training/development area. The procedure resulted in $4200$ images in the training and development set and $2800$  images in the test (accuracy assessment) set. Finally, we used $60\%$ from the training/development set for training and the rest for the development of the deep learning models.}

\begin{table}[]
	\centering
	\caption{\label{t:network_desc} The properties of the examined semantic segmentation architectures}
	\begin{tabular}{lllll}
		\textbf{Architecture} &  \textbf{Base model} & \textbf{Parameters} &  \\ \hline
           \textit{BiSeNet} &  ResNet101 &  24.75M &  \\
		\textit{SegNet} &  VGG16 &  34.97M &  \\
		\textit{Mobile U-Net} &  Not applicable & 8.87M &  \\
		\textit{DeepLabV3+} &  ResNet101 &  47.96M & \\
        \textit{FRRN-B} &  ResNet101 &  24.75M &  \\
        \textit{PSPNet} &  ResNet101 &  56M &  \\
        \textit{FC-DenseNet} &  ResNet101 &  9.27M &  \\
	\end{tabular}
\end{table}

\subsubsection{Data Augmentation}  
Further, we have employed the \textit{data augmentation} technique. The main idea behind the data augmentation is to enable improved learning by reusing original images with slight transformations such as rotation, flipping, adding Gaussian noise, or slightly changing the brightness. 
This provides additional information to the model and the dataset size is effectively increased. Moreover, an additional benefit of the data augmentation is in helping the model to learn some invariant data properties for which no examples are present in the original dataset. 
Given the sensitivity of the SAR backscatter, we did not want to augment the images in terms of the color, brightness, or by adding noise. However, we could safely employ rotations and flipping. For rotations, we only used the $90$\degree increments, giving three possible rotated versions of an image. For image flipping, we applied horizontal and vertical flipping, or both at the same time, giving another three possible versions of the original image.\footnote{Vertical flip operation switches between top-left and bottom-left image origin (reflection along the central horizontal axis), and horizontal flip switches between top-left and top-right image origin (reflection along the central vertical axis)} Notice that our images are square, so the transformations did not change the image dimensions. 
Finally, we applied the \textit{online augmentation}, as opposite to the offline version. 
In the online process, each augmented image is seen only once, and so this process yields a network that generalises better.

\subsubsection{Implementation}To apply the described semantic segmentation models, we adapted the open-source Semantic Segmentation Suite. We used Python with TensorFlow \cite{abadi2016tensorflow} backend.

\subsubsection{Hardware and Training Setup}  We trained and tested separately each of the deep learning models on a single GPU (NVIDIA GeForce GTX 1080) on a machine with 32GB of RAM.

\rev{For all the models, we used the Adam optimisation method \cite{adam_opt} with the learning rate of $0.0001$, and with the exponential decay rate for the first moment estimates of $0.9$, and for the second moment estimates of $0.999$. We applied the early stopping criterion so that, for each model, the training would automatically stop after there was no improvement in the development (validation) loss for 10 epochs. Such early stopping criteria resulted in different models being trained for a different number of epochs: from 69 (for DeepLabV3+ on the RGB SAR Ratio dataset) up to 126 (for SegNet on the RGB SAR DEM dataset) epochs. In general, all the models took slightly longer to train on the RGB SAR DEM dataset. In each case, the checkpoint for the latest model with the best result prior to stopping was saved.} Then we used that model for prediction on the test set and we report those results.

\begin{figure*}[!htb]
  	\centering
	\includegraphics[width=0.95\linewidth]{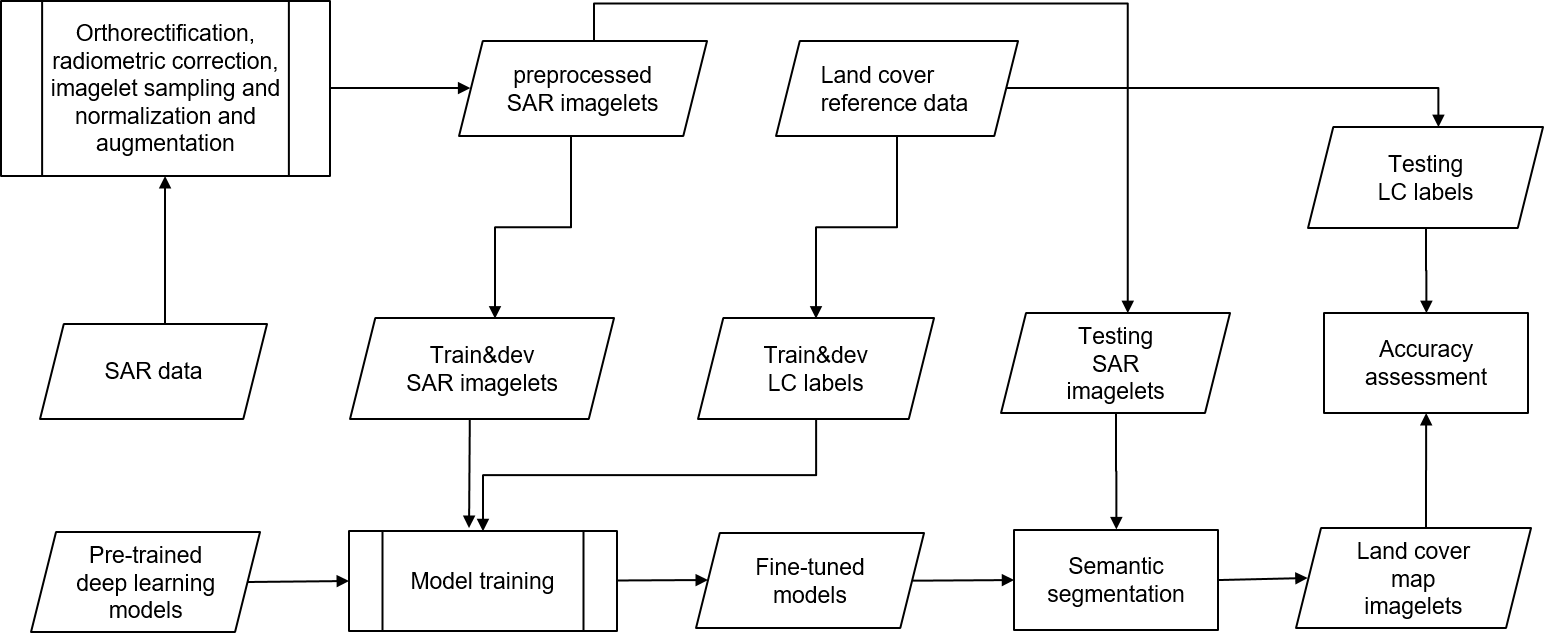}%
	\caption{\rev{General processing flowchart for RGB SAR Ratio dataset}}%
	\label{fig:flowchart}
\end{figure*}

The general processing flowchart (RGB SAR Ratio case) is shown in Figure \ref{fig:flowchart}. For the second dataset, also DEM layer is used alongside SAR data. 

\subsection{Evaluation Metrics}
In the review on the metrics used in land cover classification, Costa \textit{et al.} \cite{costa2018supervised} have found a lack of consistency, complicating intercomparison of different studies. To avoid such issues and ensure that our results are easily comparable with the literature, we thoroughly evaluated our models. For each model and class, we report the following measures of accuracy: precision, also known as producer's accuracy (PA), recall, also known as user's accuracy (UA), and overall accuracy and Kappa coefficient. 
The formulas are as follows. 

For each segmentation class (land cover type) $c$, we calculate \textit{precision (producer's accuracy)}:
$$
P_c = \frac{Tp_c}{Tp_c + Fp_c},
$$
and \textit{recall (user's accuracy)}:
$$
R_c =  \frac{Tp_c}{Tp_c + Fn_c},
$$
where $Tp_c$ represents true positive, $Fp_c$ false positive, and $Fn_c$ false negative pixels for the class $c$. 

When it comes to accuracy \cite{csurka2013good}, we calculate \textit{per class accuracy}:\footnote{Effectively, per class accuracy is defined as the recall obtained on each class.}
$$
Acc_c =  \frac{C_{ii}}{G_i},
$$
and \textit{overall pixel accuracy}:
$$
Acc_{OP} = \frac{\Sigma_{i=1}^L C_{ii}}{\Sigma_{i=1}^L G_{i}},
$$
where $C_{ij}$ is the number of pixels having a ground truth label $i$ and being classified/predicted as $j$, $G_i$ is the total number of pixels labelled with $i$, and $L$ is the number of classes. All these metrics can take values from 0 to 1.

Finally, we also use a Kappa statistic (Cohen's measure of agreement), indicating how the classification results compare to the values assigned by chance \cite{cohen1960}. Kappa statistics can take values from 0 to 1. Starting from a $k$ by $k$ confusion matrix with elements $f_{ij}$, following calculations are done:

\begin{align}
P_o &= \frac{1}{N} \sum_{j = 1}^k f_{jj}, \\
r_i &= \sum_{j = 1}^k f_{ij}, \forall i, \text{ and }
c_j = \sum_{i = 1}^k f_{ij}, \forall j, \\
P_e &= \frac{1}{N^2} \sum_{i = 1}^k r_i c_i,
\end{align}
where $P_o$ the observed proportional agreement (effectively the overall accuracy), $r_i$ and $c_j$ are the row and column totals for classes $i$ and $j$, and $P_e$ is the
expected proportion of agreement. The final measure of agreement is given by such statistic \cite{cohen1960}
\begin{align}
\kappa &= \frac{P_o - P_e}{1 - P_e}.\label{eq:kappa}
\end{align}

Depending on the value of Kappa, the observed agreement is considered as either poor (0.0 to 0.2), fair (0.2 to 0.4), moderate (0.4 to 0.6), good (0.6 to 0.8) or very good (0.8 to 1.0).

\section{Results and Discussion}
\label{sec:results}

\begin{figure*}[!htb]{}
  	\centering
  	\includegraphics[width=.80\linewidth]{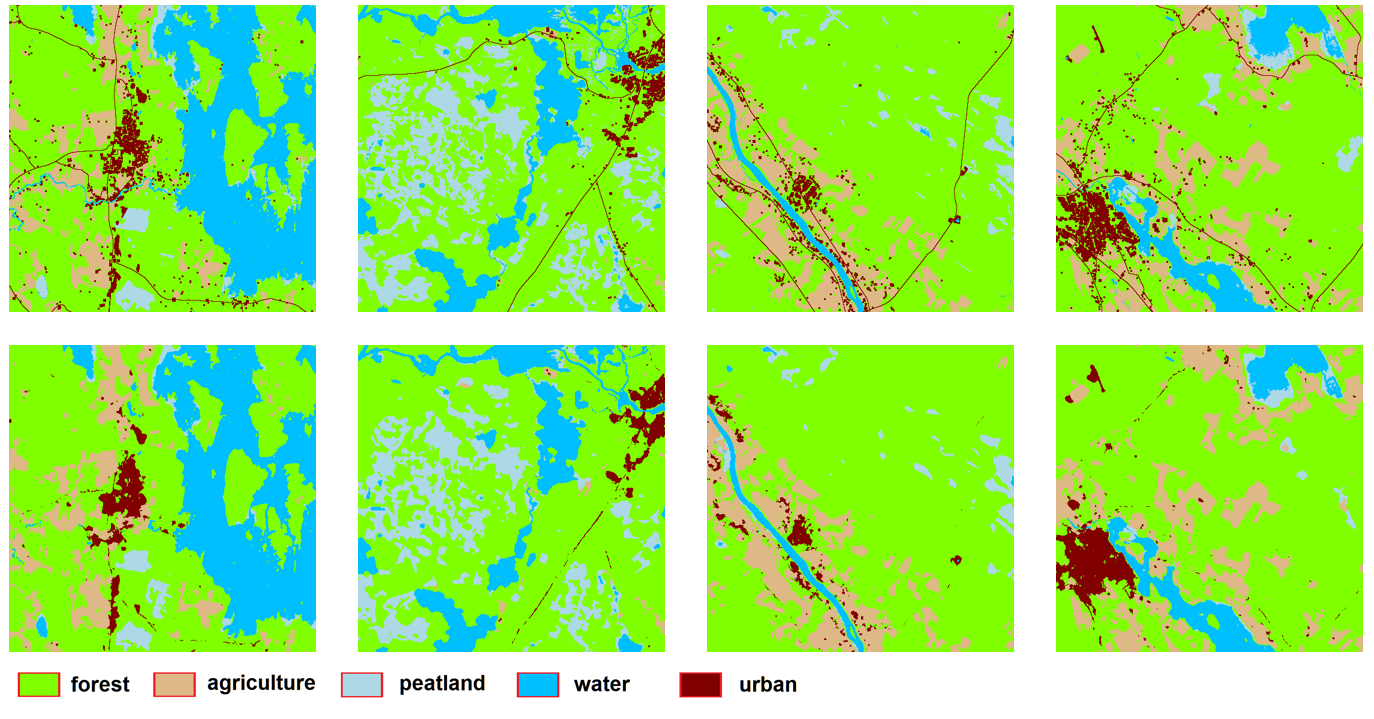}%

\caption{{Illustration of the FC-DenseNet model performance: selection of classification results, i.e., direct output of the network, without any post-processing (bottom row) versus reference CORINE based land cover (upper row).}}
\label{fig:example_res_images}
\end{figure*}

Using the experimental setup described in previous section, we evaluated the seven selected semantic segmentation models: SegNet \cite{badrinarayanan2017segnet}, PSPNet \cite{zhao2017pyramid}, BiSeNet \cite{yu2018bisenet}, DeepLabV3+ \cite{chen2018encoder,chen2018deeplab}, U-Net \cite{RFB15a,howard2017mobilenets}, FRRN-B \cite{pohlen2017full}, and FC-DenseNet \cite{jegou2017one}. The overall classification performance statistics for all studied models is gathered in Table \ref{tab:Accuracy}. Figure \ref{fig:example_res_images} shows maps produced for several imagelets with the best performing model, FC-DenseNet. Obtained results are compared to prior work and classification performance for different land cover classes is discussed further.

\begin{table*}
\caption{Summary of the classification performance and efficiency of deep learning models on the \textbf{RGB SAR Ratio} dataset (UA -- user's accuracy, PA -- producer's accuracy)}
\centering
\scalebox{0.95}{
\setlength{\tabcolsep}{0.5mm}{
\begin{tabular}{lcccccccc}
\hline
\hline
\toprule
\textbf{LC classes}  &  \textbf{Test scale ($km^{2}$)} & \multicolumn{7}{c}{\textbf{Accuracy (UA, PA \%)}}\\
& & BiSeNet & DeepLabV3+ & SegNet & FRRN-B & U-Net &  PSPNet & FC-DenseNet  \\ 
\hline
\midrule
%\textbf{RGB SAR-Ratio} \\
Urban fabric 	     & {7906}     & 59, 29   & 67, 29   &{70}, 33  &68, 33 &65, 31 & 65, 31 & 69, {34} \\
Agricultural areas   & {22659}    &63, 69   &76, 70   &79, {79}   & {80}, 77 & 69, 77 & 69, 77 & {80}, {79}   \\
Forested areas  & {199327}      &91, 96    &92, 97   &94, 97  & 94, 97 & 94, 96 & 94, 96 & 94, 97    \\
Peatland, bogs and marshes   & {10686}     &77, 48   &83, 42   &81, 61  & 77, 65 &77, 51 & 77, 51 & 78, 65     \\
Water bodies     & {53022}    &96, 86   &97, 97   &98, 98 & 98, 98 & 97, 97 & 97, 97 &98, 98     \\
\midrule
\textit{Overall Accuracy (\%)}       &          &{88.48}   &{91.37}  &{92.78}  & {92.69} & {92.25}   & {91.37} & {92.78}   \\
\textit{Kappa}                       &          &{0.758}  &{0.818}  &\textbf{0.851}   &{0.849} &{0.839} & {0.823} & \textbf{0.851}    \\
\midrule
\textit{Average inference time (s / image)}    &          & {0.043 }  &\textbf{0.031}   & {0.073}   & { 0.143} & {0.085}   & {0.053} & {0.196} \\
\hline
\hline
\end{tabular}
}}
\label{tab:Accuracy}
\end{table*}

\begin{table*}
\caption{Summary of the classification performance and efficiency of deep learning models on the \textbf{RGB SAR DEM} dataset (UA -- user's accuracy, PA -- producer's accuracy)}
\centering
\scalebox{0.95}{
\setlength{\tabcolsep}{0.5mm}{
\begin{tabular}{lcccccccc}
\hline
\hline
\toprule
\textbf{LC classes}  &  \textbf{Test scale ($km^{2}$)} & \multicolumn{7}{c}{\textbf{Accuracy (UA, PA \%)}}\\
& & BiSeNet & DeepLabV3+ & SegNet & FRRN-B & U-Net &  PSPNet & FC-DenseNet  \\ 
\hline
\midrule
%\textbf{RGB SAR-Ratio} \\
Urban fabric 	     & 7906     &50, 25   &52, 26   &66, 39  & 66, 38 &65, 34 & 62, 30 & 69, 35 \\
Agricultural areas   & 22659    &63, 65   &70,55   &79, 82   & 77, 83 & 78, 77 & 70, 72 & 80, 80   \\
Forested areas  & 199327      &90, 95    &90,97   &95, 97  & 95, 96 & 94, 97 & 92, 97 & 94, 97    \\
Peatland, bogs and marshes   & 20990     &75, 34   &{78, 31}   &77, 67  & 78, 65 & 77, 63 & 77, 47 & 79, 64     \\
Water bodies     & 53022    &94, 91   &94, 97   &98, 98 & 98, 98 &98, 97 & 97, 94 &98, 98     \\
\midrule
\textit{Overall Accuracy (\%)}       &          & {87.86}   &{89.22}  & {93.05}   & {92.90} & {92.53}   & {90.48} & {93.01}   \\
\textit{Kappa}                       &          &{0.745}  &{0.770}  &{0.858}   &{0.855} &{0.846} & {0.801} & {0.856}    \\
\midrule
\textit{Average inference time (s / image)}  &         &{0.045}  & {0.039}  & {0.074}   & {0.143} & {0.086} & {0.053} & {0.192}    \\
\hline
\hline
\end{tabular}
}}
\label{tab:Accuracy}
\end{table*}

\begin{table}[]
\caption{Confusion matrix for classification with FC-DenseNet model for RGB SAR Ratio dataset.}
\scalebox{.9}{
\setlength{\tabcolsep}{.5mm}{
\begin{tabular}{cccccccc}
\hline
\multicolumn{8}{c}{FC-DenseNet103}                                                                                            \\ \hline
LC class & \multicolumn{5}{c}{Sentinel-1 class}                                        & \multicolumn{2}{l}{}                  \\ \hline
        & urban        & agriculture         & forest         & peatland        & \multicolumn{1}{c|}{water}        & \multicolumn{1}{c|}{total}     & PA   \\ \hline
\multicolumn{1}{l|}{urban}       & 6674628  & 2567434    & 10281910  & 127574  & \multicolumn{1}{c|}{114168}   & \multicolumn{1}{c|}{19765714}  & 33.8 \\
\multicolumn{1}{l|}{agriculture}       & 406663    & 44490302 & 11169843   & 514909   & \multicolumn{1}{c|}{66019}  & \multicolumn{1}{c|}{56647736} & 78.5 \\
\multicolumn{1}{l|}{forest}       & 2485460  & 7729542   & 483087561 & 3583613 & \multicolumn{1}{c|}{1432249}  & \multicolumn{1}{c|}{498318425} & 96.9 \\
\multicolumn{1}{l|}{peatland}       & 47057   & 844064    & 7010103  & 17231477 & \multicolumn{1}{c|}{1583214}   & \multicolumn{1}{c|}{26715915}  & 64.5 \\
\multicolumn{1}{l|}{water}       & 84660    & 221139   & 2159293  & 594050  & \multicolumn{1}{c|}{129496268} & \multicolumn{1}{c|}{132555410}  & 97.7 \\ \hline
\multicolumn{1}{l|}{total}   & 9698464 & 55852481 & 513708710 & 22051623 & \multicolumn{1}{c|}{132691918} & \multicolumn{1}{c|}{73400320} &      \\ \hline
\multicolumn{1}{l|}{UA}      & 68.8     & 79.7      & 94.0      & 78.1     & \multicolumn{1}{c|}{97.6}     & \multicolumn{1}{l|}{}          & 92.8 \\ \hline
\end{tabular}
}}
\end{table}

\begin{figure*}
     \centering
     \begin{subfigure}[b]{0.7\textwidth}
         \centering
          \includegraphics[width=.44\linewidth]{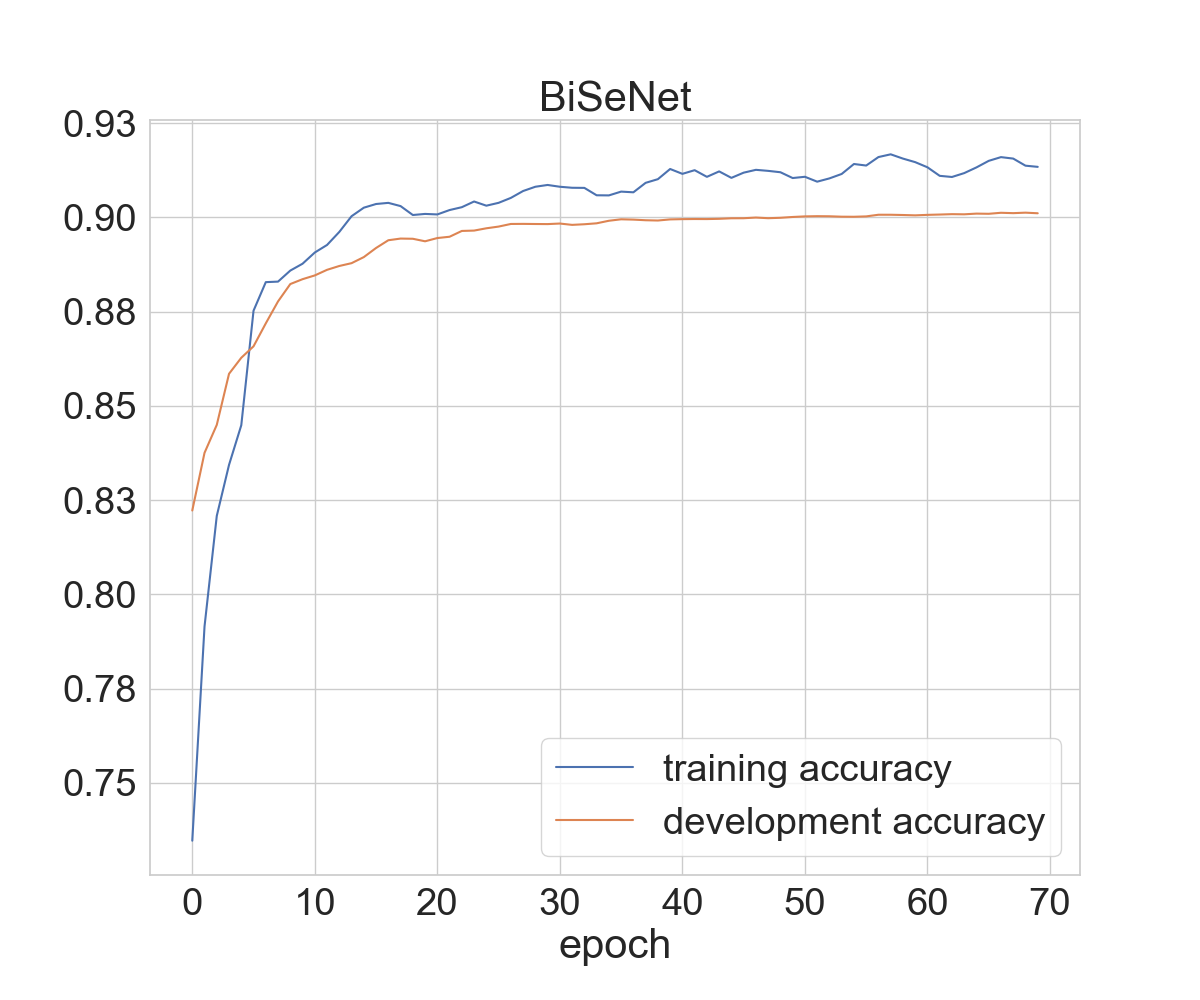}
          \includegraphics[width=.44\linewidth]{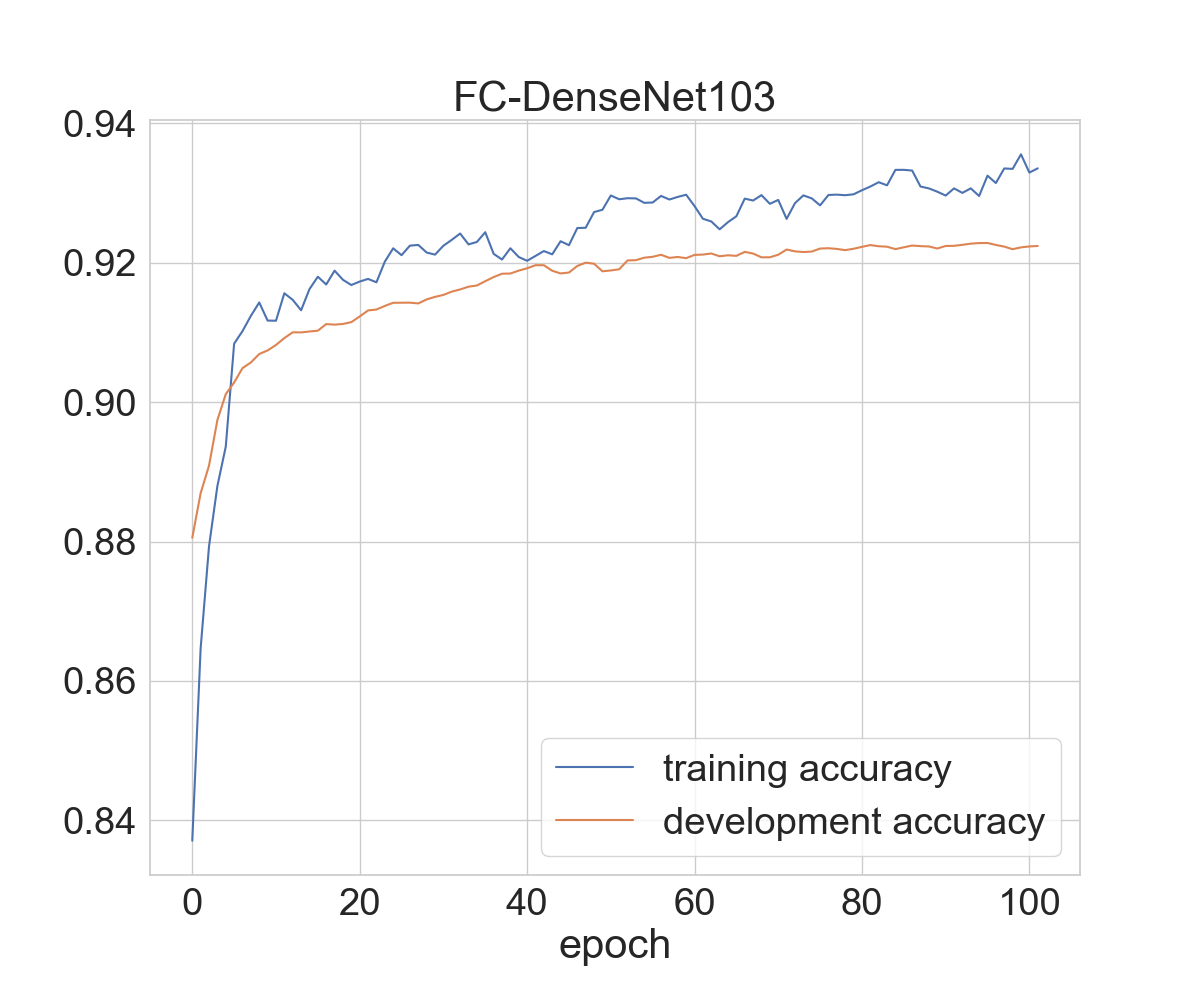}
         \caption{RGB SAR Ratio dataset}
         \label{fig:acc_nodeDEM}
     \end{subfigure}
     \hfill
     \begin{subfigure}[b]{0.7\textwidth}
         \centering
  	      \includegraphics[width=.44\linewidth]{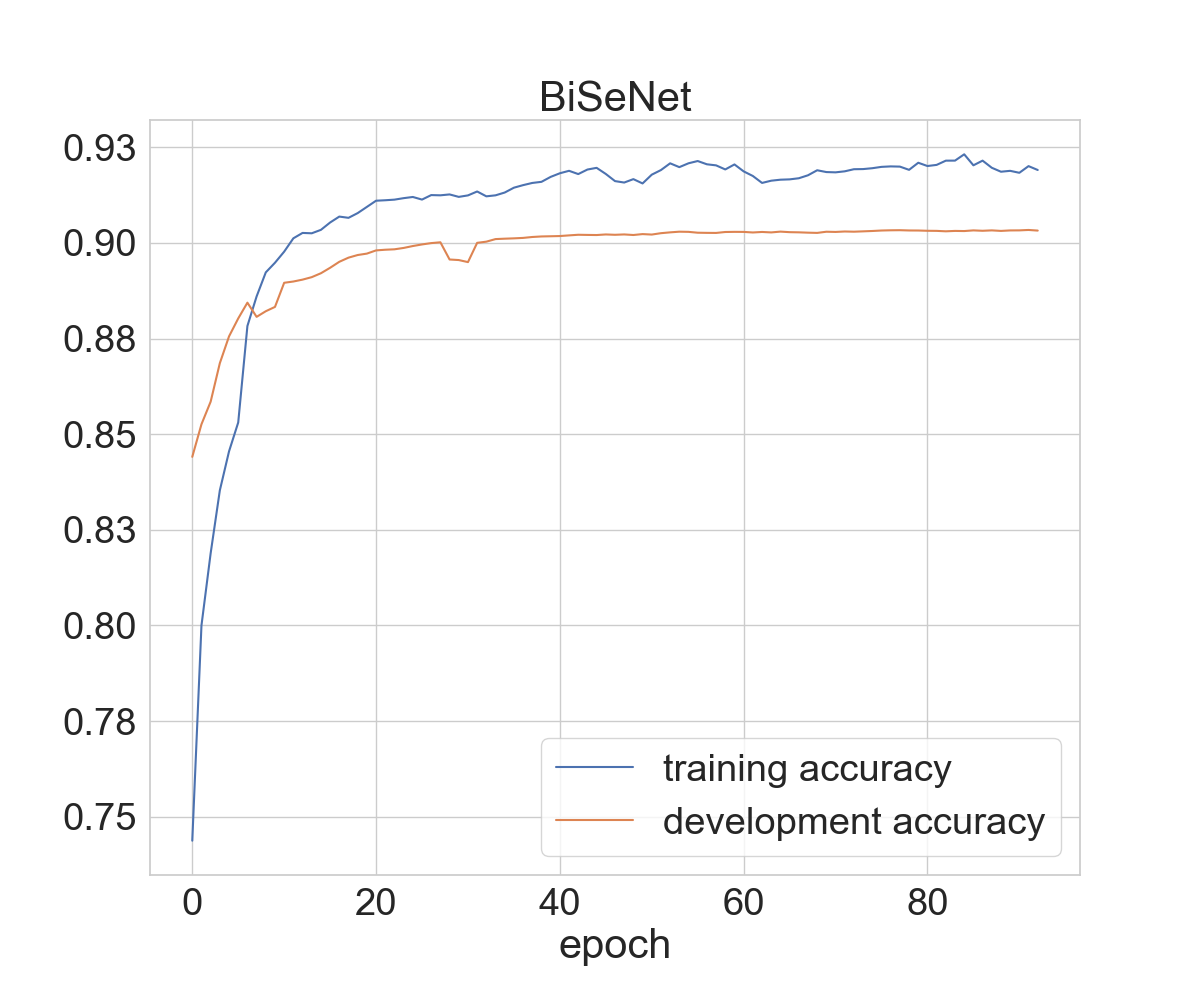}%
          \includegraphics[width=.44\linewidth]{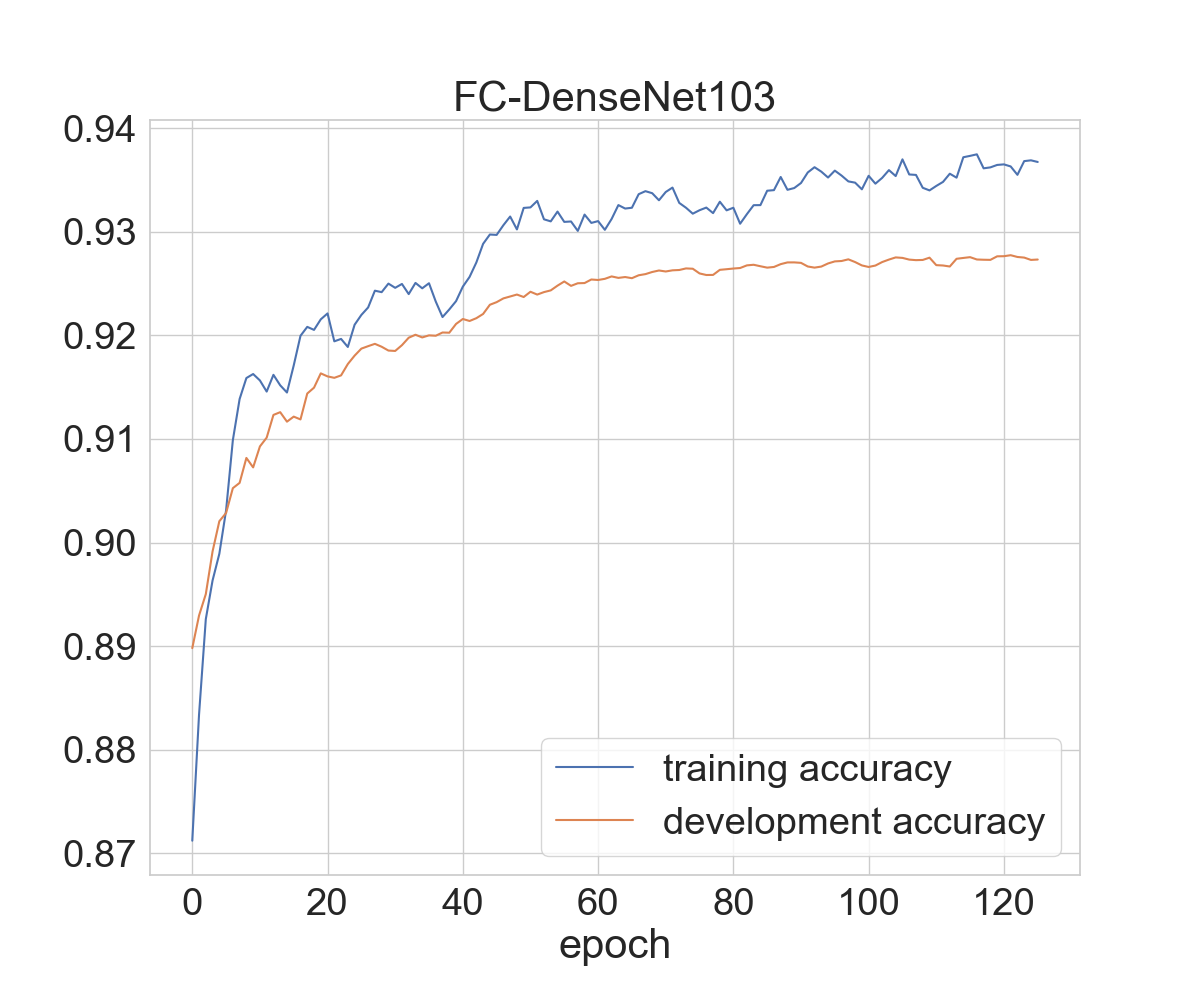}
  	      \caption{RGB SAR DEM dataset}
         \label{fig:acc_DEM}
     \end{subfigure}
    \caption{\rev{Accuracy curves during training and development on both datasets for the fastest (BiSeNet) and the slowest (FC-DenseNet) model. The early-stopping criteria with 10 epochs of no improvement for development loss was applied.}}
    \label{fig:accur_curves}
\end{figure*}

\subsection{Classification Performance} 
%\textcolor{red}{SANJA: Let's tell something about the accuracy curves and how well the Early-stopping criterion has worked for our experiments. Do inference times make sense if we use Early stopping?}

%\textcolor{red}{Commenting about DEM vs no-DEM accuracy with one of quicker models -BiSeNet }

% All the models performed relatively well in terms of classification, achieving the overall accuracy above $83\%$. Three models performed particularly well, achieving the accuracy score above $89\%$: SegNet, FRRN-B, and the best performing model FC-DenseNet, which achieved the accuracy of $90.7\%$. 

\rev{
All the models performed relatively well on both datasets achieving the overall accuracy above $87\%$ for each model. Four models performed particularly well, achieving the accuracy score above $92\%$ on both datasets; those are: FRRN-B, U-Net, SegNet, and FC-DenseNet. The two latter models were also somewhat better than others in terms of kappa statistics, and, along with FRRN-B, were the best models also with respect to class-wise user's and producer's accuracy.  %, i.e., SegNet and FC-DenseNet share the best accuracy scores on both datasets, 92\% and 93\%, on RGB SAR Ratio and RGB SAR DEM, respectively. 
The advantage for SegNet is that its training and inference times were 2.5 better compared to the FC-DenseNet of similar accuracy. BiSeNet and DeepLabV3 were performing somewhat worse than other five models particularly in terms of agreement (kappa was 0.75-0.82), but also overall accuracy was lower, most strongly for BiSeNet. Overall accuracy and class-wise accuracies obtained on completely independent test dataset were still remarkably high compared to other reported results in the literature when more conventional statistical or traditional machine learning approaches were used with C-band SAR data \cite{balzter2015corine, thiel2009}. Further in-depth comparison can be found in Section IV-C.}

Before further analysis, let us recall that CORINE is not exclusively a land cover map, but rather land cover and land use map, thus some specific classes can differ from ecological classes observed by Sentinel-1. Also, the aggregation to CLC Level-1 is sometimes not strictly "ecological" or complies to physics surface scattering considerations. For example, airports, major industrial areas and road network often exhibit areas similar to field, presence of trees and green vegetation near summer cottages can cause them exhibit signatures close to forest rather than urban, sometimes forest on the rocky terrain can be misclassified as urban instead due to presence of very bright targets and strong disruptive features, while confusion between peatland and agricultural and grassland areas is also common. Finally, the accuracy of the CORINE data is only somewhat higher than 90\%.

As for the results across the different land classes, all the models performed particularly well in recognising the water bodies and forested areas, while the urban fabric represented the most challenging class for all the models.  %We expect that the inclusion of the DEM as one layer in the training images has helped to achieve good results on the water bodies class for most of the models (except for BiSeNet, all the models achieved both the user and producer accuracy above $90\%$).
The urban class was particularly challenging for the following main reasons. First, this is still esentially a land use class, with continuous urban fabric (easy to recognize by radar) representing only a moderate fraction of the whole class. It also changes the most, as new houses, roads, and urban areas are built. %While we took the most suitable available CORINE class in terms of time for our Sentinel-1 images, there are almost certain differences between the urban class as it was in 2012 and in 2015-2016. 
Second, the CORINE map itself does not have a perfect accuracy, neither aggregation rules are perfect. As a matter of fact, in majority of studies where SAR based classification was done versus CLC or similar data, a poor or modest overall agreement was observed for urban land use areas \cite{lonnqvist2010, lumsdon2005, antropov2012, antropov2014}, while the user's accuracy was strongly  higher than producer's \cite{antropov2011}. The latter is exactly due to radar being able to sense sharp boundaries and bright targets very well whereas such bright targets often don't dominate the whole urban land-use class. Importantly, relatively good performance was obtained in mapping agricultural and wetland areas, particularly well differentiating between them while this is often problematic with other remote sensing instruments.
%\oleg{We argue that any inaccuracies present will be particularly attenuated in our models for the urban class because of the sharp and sudden boundary changes in this class, unlike for the others, such as forest and water. The top performing model, i.e., FC-DenseNet, performed the best across all the classes. It is particularly notable that it achieved the user accuracy, i.e., precision for the urban class of $62\%$, improving on it significantly compared to all the other models. Nevertheless, its score on the producer accuracy, i.e., recall on this class of $27\%$ is outperformed by the two other top models, i.e., SegNet and FRRN-B.}

%\sanja{Oleg, please double-check the text below now. I added a new description on both datasets in the dataset section, so we should refer to them with those names: RGB SAR Corss-Pol and RGB SAR DEM now.}

{In addition to the VH-to-VV ratio, we have tested topographic DEM as a candidate for the third layer in RGB imagelets. However, the difference in classification accuracy (summarized in Tables 4 and 5) was marginal. This limited gain in accuracy can be explained by several reasons. Firstly, accuracies achieved using only SAR data were large overall, and relatively large for the majority of classes, particularly water. Additionally, DEM variation in the study area was limited, mostly within 0-300 meters asl. If the DEM variation was higher, this could affect land use and vegetation, and result in a larger impact of DEM on classification accuracy (e.g., in Scotland, Norway and many other countries). Moreover, the DEM used in the study is essentially a topographic digital terrain model and doesn't include forest canopy height models or high-contrast features of urban structures within settlements, which could potentially boost the classification accuracy for urban and forest classes. }

%If the study area included larger variation in DEM heights with high hills and mountains, DEM layer would be instrumental for separating particularly classes where scattering characteristics are similar, but land use is different, such as e.g. some agricultural fields and high-land grassland and pastures. However, in the context of this study we used CLC level-1 classes where the all these smaller LULC classes belong to one class. Keeping in mind low variation of DEM in southern and central Finland, the whole testing was a bit strange idea in the first place but we proceeded with this. As the results were ready and calculated, we wanted them to be published (not to mention one of the reviewers was asking), thus we included them to Table X. In future, we plan to test also possible  accuracies with smaller classes - CLC level-2 and 3, where benefit DEM can be perhaps more useful. However, there overall accuracies might be too low to get it published somewhere - unless we will use multitemporal signatures of SAR data.  

We mentioned the issues of SAR backscattering sensitivity to several ground factors so that the same classes might appear differently on the images between countries or between distant areas within a country. An interesting indication of our study, however, is that the deep learning models might be able to deal with this issue. Namely, we used the models pre-trained on ImageNet and fine tuned them with a relatively small number of Sentinel-1 images. The models learned to recognize varying types of the backscattering signal across the country of Finland. This indicates that with a similar type of fine-tuning, present models could be relatively easily adapted to the other areas and countries, with different SAR backscattering patterns. Such robustness and adaptability of the deep learning models come from their automatic learning of feature representation, without the need for a human expert pre-defining those features.

\subsection{Computational Performance}
\rev{ The training times with our hardware configuration took from $1$ up to $2.5$ days for the different models. This could be significantly improved by training each model using a multi-GPU system instead of a single-GPU in our experiments.}

In terms of the inference time, we also saw the differences in the performance. In Table \ref{tab:Accuracy}, we present the average inference time per $512\ px\times 512\ px$  imagelet that we worked with. The results show that there is a trade-off between classification and computational performance: the best models in terms of classification results (i.e., FC-DenseNet and FRRN-B) take several times longer inference time compared to the rest. \rev{A positive exception in this regard is the SegNet model, which achieved the best classification results together with FC-DenseNet but with 2.5 times better inference time.} Depending on the application, this might or might not be of particular importance.

\subsection{Comparison to Similar Work}

Obtained results compare favourably to previous similar studies on land cover classification with SAR data \cite{antropov2012,antropov2014,lonnqvist2010,lumsdon2005,laurin2013optical,balzter2015}. Depending on the level of classes aggregation (4-5 major classes or more), with using mostly statistical or classical machine learning approaches reported classification accuracies were as high as 80-87\% to as low as 30\% when only SAR imagery were used.

Two recent studies that employed neural networks to SAR imagery classification (albeit in combination with satellite optical data) for land cover mapping were \cite{laurin2013optical} and \cite{kussul2017deep}, with reported classification accuracies of up to 97.5\% and $94.6$\%, respectively. 

The best models in our experiments achieved the overall accuracy of $93\%$. However, our results are obtained using solely the SAR imagery. In contrast,  SAR imagery (PALSAR) alone yielded the overall accuracy of 78.1\% in \cite{laurin2013optical}. The types of classes they studied are also different compared to ours (crops versus vegetation versus land cover types) and our study is performed on a larger area. Importantly, the previous studies have applied different types of models (regular NNs versus CNN versus semantic segmentation). In particular, the CNN models work on the $7 \times 7$ resolution windows, while we have applied more advanced semantic segmentation models, which work on the level of a pixel.  Keeping in mind findings from \cite{laurin2013optical} that the addition of optical images on top of SAR improved the results for over 10\%, we expect that our models would perform comparably well or outperform these previous works if applied to a combined SAR and optical imagery. 

In terms of the deep learning setup, the most similar to ours are the studies \cite{mahdianpari2018very} and \cite{mohammadimanesh2019}. However, RapidEye optical imagery at 5 m spatial resolution was used in \cite{mahdianpari2018very}, and the test site was considerably smaller. Study \cite{mohammadimanesh2019}, similar to our research, relied exclusively on SAR imagery, however, fully polarimetric images, and acquired by RADARSAT-2 at considerably better resolution. They have developed an FCN-type of a semantic segmentation model \textit{`specifically designed for the classification of wetland complexes using PolSAR imagery'}.  Using this model to classify eight wetland map classes, they achieved the overall accuracy of $93$\%.  However, because their model is designed specifically for wetland complexes, it is not clear if such a model would generalize to other types of areas. Compared to our study, they have focused on a considerably smaller area (nearly the size of a single imagelet we used), and on a very specific task (wetland types mapping). Thus, it is not readily clear how general their approach is and how it compares to our presented approach.

\subsection{Outlook and Future Work}
There are several lines for potential improvement based on the results of this study, as well as future work directions. 

First, using even a larger set of Sentinel-1 images can be recommended since for the supervised deep learning models large amounts of data are crucial.  Here, we processed only $6888$ imagelets altogether, but deep learning algorithms become efficient typically only once they are  trained with hundreds of thousands or millions of images. 

Second, if SAR images and reference data of a higher resolution are used, we expect better classification performance, too, as smaller details could be potentially captured. Also, better agreement in acquisition timing of reference and SAR imagery can be recommended. The reference and training data should come from the same months or year if possible, and that the reference maps should represent the reality as accurately as possible. The models in our experiments were certainly limited by the CORINE's own limited accuracy. 

Third, in this study we have tested the effectiveness of off-the-shelf deep learning models for land cover mapping from SAR data. While the results show their effectiveness, it is also likely that the novel types of models, specifically developed for the radar data (such as \cite{mohammadimanesh2019}), will yield even better results. Based on our results, we suggest DenseNet- and SegNet-based models as a starting point. In particular, one could develop the deep learning models to handle directly the SLC data which preserve the phase information.

Focusing on a single season is both an advantage and a limitation. Importantly, we have avoided confusion between SAR signatures varying seasonally for several land cover classes. However, multitemporal dynamics itself can be potentially used as an additional useful class-discriminating parameter. Incorporating seasonal dynamics of each land cover pixel (as a time series) is left for future work, perhaps with additional need to incorporate recurrent neural networks into the approach.

As discussed in Section 3.1.1, it could be suitable to use more detailed (specific) land cover classes, as aggregation of smaller LC classes into CORINE super-classes is not exactly ecological, leading to mixing several distinct SAR signatures in one class, and thus causing additional confusion for the classifier. Later, classified specific classes can be aggregated into larger classes, potentially showing improved performance \cite{hame2013}.

Finally, we have used only SAR images and a freely-available DEM model for the presented large-scale land cover mapping. If one were to combine other type of remote sensing images, in particular the optical images, we expect that the results would significantly improve. This is true for those areas where such imagery can be collected due to cloud coverage, while in operational scenario  it would potentially require use of at least two models (with and without optical satellite imagery). It is also important to access added value of SAR imagery with deep learning models when optical satellite images are available, as well as possible data fusion and decision fusion scenarios, before a  decision on the mapping approach is done \cite{hame2013}.

\section{Conclusion}
\rev{Our study demonstrated the potential for applying state-of-the-art semantic segmentation models to SAR image classification with high accuracy.  Several models were benchmarked in a countrywide classification experiment using Sentinel-1 IW-mode SAR data, reaching nearly 93\% overall classification accuracy with the best performing models (SegNet  and FC-DenseNet). This indicates strong potential for using pre-trained CNNs for further fine-tuning and seems particularly suitable when the number of training images is limited (to thousand or tens of thousands instead of millions). In addition to suggesting the best candidate semantic segmentation models for land cover mapping with SAR data (that is, the DenseNet-based models), our study offers baseline results against which the newly proposed models should be evaluated. Several possible improvements for future work were identified, including the necessity for testing multitemporal approaches, data fusion, and very high-resolution SAR imagery, as well as developing models specifically for SAR.}

\section*{Acknowledgment}
The authors were supported by ICEYE Oy during the study. S\v{S} was also supported by EIT Digital and OA was also supported by Aalto university and VTT. Authors thank reviewers for careful reading of the manuscript and their valuable comments.

\section*{Data availability}
\rev{The implementation scripts with documentation are available on GitHub,\footnote{\url{https://github.com/sanja7s/DL\_SemSAR\_with\_docs}} the original Sentinel-1 images can be downloaded from SciHub,\footnote{\url{https://scihub.copernicus.eu}} and the processed train/development and test data are published on Zenodo\footnote{\url{https://zenodo.org}} and IEEE DataPort.} 
%\textcolor{red}{actually only original Sentinel-1 images are available on Scihub. Sanja, could you please change this to the name of the portal where we plan to publish the preprocessed data?}

\ifCLASSOPTIONcaptionsoff
  \newpage
\fi



\bibliographystyle{IEEEtran}
\bibliography{jstars_bib.bib}

\end{document}